\documentclass{raa}
\usepackage{graphicx,times}
\usepackage{natbib}
\usepackage{amssymb,amsmath}

\bibpunct{(}{)}{;}{a}{}{,}

\newcommand{\velunits}{~$\rm km~s^{-1}$}

\newcommand{\vph}{\ensuremath{v_{\mathrm{ph}}}}

\def\vec#1{\ensuremath{\mathbf{#1}}}

\begin{document}

   \title{Period ratios for standing kink and sausage modes in magnetized structures with siphon flow on the Sun}

 \volnopage{ {\bf 2012} Vol.\ {\bf X} No. {\bf XX}, 000--000}
   \setcounter{page}{1}

   \author{Hui Yu\inst{1}, Shao-Xia Chen\inst{1}, Bo Li\inst{1}, \& Li-Dong Xia\inst{1}}

   \institute{
   Shandong Provincial Key Laboratory of Optical Astronomy and
Solar-Terrestrial Environment, Institute of Space Sciences, Shandong University, Weihai, 264209, China; {\it ruxanna@sdu.edu.cn}}
	

\vs \no
  \date{Received~~2015 month day; accepted~~2015~~month day}

\abstract{Standing oscillations with multiple periods were found in a number of atmospheric structures on the Sun. The ratio of the period of the fundamental to twice the one of its first overtone, $P_1/2P_2$, is important in applications of solar magneto-seismology. We examine how field-aligned flows impact $P_1/2P_2$ of standing modes in solar magnetic cylinders. For coronal loops, the flow effects are significant for both fast kink and sausage modes. For kink ones, they reduce $P_1/2P_2$ by up to 17\% relative to the static case even when the density contrast between the loop and its surroundings approaches infinity. For sausage modes, the reduction in $P_1/2P_2$ due to flow is typically $\lesssim 5.5\%$ compared with the static case. However, the threshold aspect ratio, only above which can trapped sausage modes be supported, may increase dramatically with the flow magnitude. For photospheric tubes, the flow effect on $P_1/2P_2$ is not as strong. However, when applied to sausage modes, introducing field-aligned flows offers more possibilities in interpreting the multiple periods recently measured. We conclude that field-aligned flows should be taken into account to help better understand what causes the departure of $P_1/2P_2$ from unity.
}

   \authorrunning{Hui Yu et al. }            
   \titlerunning{Period ratios for standing kink and sausage modes}  
   \maketitle

%
\section{Introduction}           
\label{sect:intro}

The frequently measured waves and oscillations can be exploited to deduce the physical parameters of the structured solar atmosphere
   that are otherwise difficult to yield, thanks to the diagnostic power of the solar magneto-seismology (SMS)
   \citep[see e.g., the reviews by][]{2000SoPh..193..139R, 2004psci.book.....A,2005LRSP....2....3N,2009SSRv..149....1N, 2011SSRv..158..167E,2012RSPTA.370.3193D}.
In the context of SMS, multiple periodicities interpreted as a fundamental standing mode and its overtones
    detected in a substantial number of oscillating structures
    are playing an increasingly important role~\citep[see][for recent reviews]{2009SSRv..149....3A, 2009SSRv..149..199R}.
In the case of standing kink oscillations,
   both two~\citep[e.g.,][]{2004SoPh..223...77V,2007A&A...473..959V}
   and three periodicities~\citep{2007ApJ...664.1210D, 2009A&A...508.1485V,2009A&A...493..259I,2013SoPh..284..559K}
   have been found.
Moreover, the ratio between the period of the fundamental and twice the period of its first overtone, $P_1/2P_2$, deviates in general from unity.
This was first found by~\citet{2004SoPh..223...77V} in two loops in a post-flare arcade observed by the Transition Region
   and Corona Explorer ($TRACE$) in its 171\AA\ passband on 2001 April 15,
   where values of 0.91 and 0.82 were measured for $P_1/2P_2$.
A similar value ($0.9$) was found for $TRACE$ 171\AA\ loops on 1998 November 23~\citep{2007A&A...473..959V},
   and also in flaring loops as measured with the Nobeyama Radioheliograph ($NoRH$) on 2002 July 3
   where $P_1/2P_2$ is deduced to be $0.83$~\citep{2013SoPh..284..559K}.
In this latter study the deviation $P_1/2P_2$ from unity is likely to be associated with wave dispersion at a finite aspect ratio of
   the flaring loop.
However, loops seem thin in EUV images, thereby prompting~\citet{2005ApJ...624L..57A} to attribute the finite $1-P_1/2P_2$
   to the longitudinal structuring in loop densities, given that wave dispersion is expected to be minimal
    for kink modes supported by a static longitudinally uniform loop with tiny aspect ratios.
When it comes to standing sausage modes, fundamental or global modes together with their first overtones were identified.
In flare-associated quasi-periodic pulsations measured with $NoRH$ on 2000 January 12, the global (fundamental) sausage mode
   was found to have a $P_1$ of $14-17$~s, and its first overtone
   corresponds to a $P_2$ of $8-11$~s~\citep{2003A&A...412L...7N, 2005A&A...439..727M}.
Interestingly, sausage modes were also seen in cool post-flare loops in high spatial resolution H$_\alpha$ images
   and correspond to a $P_1 \approx 587$~s and a $P_2 \approx 349$~s~\citep{2008MNRAS.388.1899S}.
Actually, sausage modes have been directly imaged in magnetic pores~\citep{2011ApJ...729L..18M}
   and chromosphere~\citep{2012NatCom..3.1315M}.
Using the Rapid Oscillations in the Solar Atmosphere ($ROSA$) instrument situated at the Dunn Solar Telescope,
    the former study employed an Empirical Mode Decomposition (EMD) analysis to reveal a number of periods,
    some of which seem to correspond to a fundamental mode and its higher overtones
    with the standing mode set up by reflections between the photosphere and transition region.

In the solar atmosphere, flows seem ubiquitous~\citep[e.g.,][]{2004psci.book.....A},
   and have been found in oscillating structures in particular~\citep[e.g.,][]{2008A&A...482L...9O, 2008MNRAS.388.1899S}.
In the coronal case, where the flow speeds tend to be $\lesssim 100$~\velunits\ and therefore
   well below the Alfv\'{e}n speed, they are not necessarily always small.
As a matter of fact, speeds reaching the Alfv\'enic range ($\sim 10^3$\velunits)
   have been seen associated with explosive events~\citep[e.g.,][]{2003SoPh..217..267I,2005A&A...438.1099H}.
In the context of standing modes supported by loops, a siphon flow causes their phases to depend on the locations, which is true
   even for the fundamental mode where only two permanent nodes are present and are located at loop footpoints.
Actually this location-dependent phase distribution was seen for the standing kink mode measured with $TRACE$ and $SOHO$
   on 2001 September 15 ~\citep{2010ApJ...717..458V}, and yields a flow speed indeed in the Alfv\'enic regime~\citep{2011ApJ...729L..22T}.
The authors went on to find that neglecting the flows leads to an underestimation
   of the loop magnetic field strength by a factor of three.

Given that multiple periodicities have received considerable interest,
   and that a significant flow may play an important role as far as the applications of the solar magneto-seismology are concerned,
   one naturally asks: how do the flows affect multiple periodicities from a theoretical perspective?
   and what would be the observational implications?
In a slab geometry, these questions were addressed by~\citet{2013ApJ...767..169L} (hereafter paper I) where a rather comprehensive
   analytical and numerical examination was conducted.
In cylindrical geometry, the flow effect on the period ratio for standing kink modes
   was assessed by~\citet{2010SoPh..267..377R} for thin coronal loops.
The present work extends both paper I and the one by~\citet{2010SoPh..267..377R}
    by examining how the flows affect the dispersion properties and hence the period ratios
    of both standing kink and sausage modes supported by a magnetized cylinder of arbitrary aspect ratio.
Besides, in addition to a coronal environment, a photospheric one will also be examined in detail to
    demonstrate how introducing a flow helps offer more possibilities in interpreting the recently
    measured multiple periods in oscillating photospheric structures.

This paper is organized as follows.
Section~\ref{sec_DR_Oview} presents a brief description of the cylinder dispersion relation.
Section~\ref{sec_Cor_Pratio} is concerned with coronal cylinders, where we first
   give an overview of the dispersion diagrams,
   briefly describe a graphical means to compute the period ratios,
   and examine how the flow affects the period ratios for standing kink and sausage modes.
Likewise, section~\ref{sec_Pho_Pratio} examines in detail isolated photospheric cylinders.
Finally, section~\ref{sec_summary} summarizes the results, ending with some
   concluding remarks.

\section{Cylinder Dispersion Relation}
\label{sec_DR_Oview}
\begin{figure}
\centerline{\includegraphics[width=0.6\columnwidth]{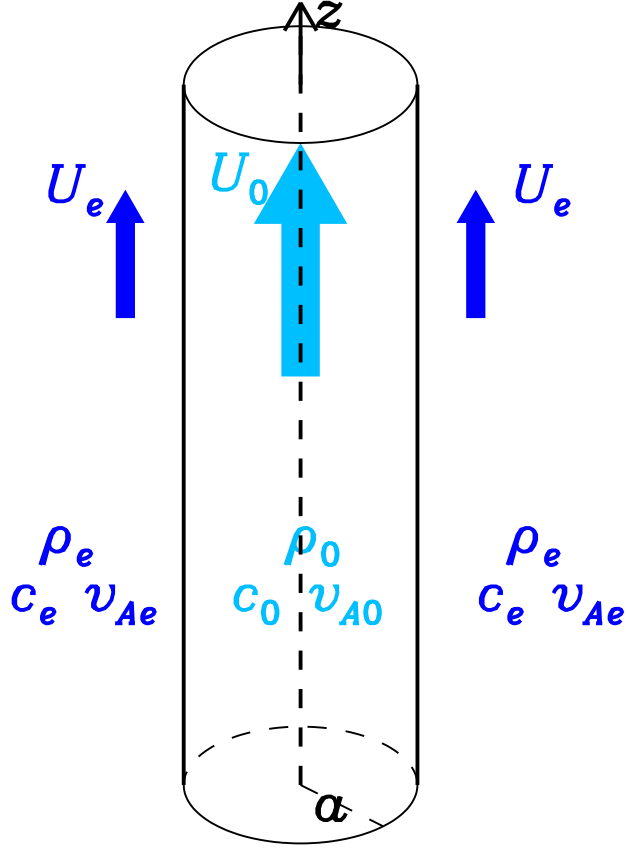}}
 \caption{Schematic diagram illustrating the magnetic cylinder (denoted by subscript $0$)
    and its environment (subscript ${\rm e}$).
 The variables $\rho_i, c_i, v_{{\rm A}i}$ and $U_i$ ($i=0, {\rm e}$) represent the mass density,
    adiabatic sound speed, Alfv\'en speed, and the flow speed, respectively.
}
 \label{fig_cyl_illus}
\end{figure}
Consider a cylinder of radius $a$ with time-independent field-aligned flows.
As illustrated in Fig.\ref{fig_cyl_illus},
    the cylinder is infinite in the $z$-direction,
    and is bordered by the interface $r=a$ in a cylindrical coordinate system $(r, \theta, z)$.
The physical parameters take the form of a step function, characterized by
    their values external to (denoted by a subscript ${\rm e}$)
    and inside (subscript $0$) the cylinder.
The background magnetic fields ($\vec{B}_0$ and $\vec{B}_{\rm e}$), together with the
    flow velocities ($\vec{U}_0$
    and $\vec{U}_{\rm e}$), are in the $z$-direction.
Let $\rho$ and $p$ denote the mass density and thermal pressure.
It follows from the force balance condition across the interface that
\begin{eqnarray}
\frac{\rho_{\rm e}}{\rho_0} = \frac{2 c_0^2 + \gamma v_{{\rm A}0}^2}{2 c_{\rm e}^2 + \gamma v_{\rm Ae}^2} ,
\label{eq_rhoe0}
\end{eqnarray}
    where $\gamma=5/3$ is the adiabatic index,
    $c=\sqrt{\gamma p/\rho}$ is the adiabatic sound
    and $v_{\rm A}=\sqrt{B^2/4\pi\rho}$ is the Alfv\'{e}n speed.
It is also necessary to introduce the tube speeds, $c_{{\rm T}i}$ ($i=0, {\rm e}$),
\begin{eqnarray}
 c_{{\rm T}i}^2 = \frac{c_i^2 v_{{\rm A} i}^2}{c_i^2 + v_{{\rm A} i}^2} ,
 \label{eq_def_ct}
\end{eqnarray}
    and the kink speed $c_k$,
\begin{eqnarray}
 c_k^2 = \hat{\rho}_0 v_{{\rm A}0}^2 + \hat{\rho}_{\rm e} v_{\rm Ae}^2,
 \label{eq_def_ck}
\end{eqnarray}
    where $\hat{\rho}_i = \rho_i/(\rho_0 + \rho_{\rm e})$ is the fractional density with $i=0, {\rm e}$.

The dispersion relation (DR) for linear waves trapped in a cylinder with flow has been examined
    by a number of authors~\citep[e.g.,][]{1991PPCF...33..333N, 1999PPCF...41.1421S, 2003SoPh..217..199T,2009A&A...498L..29V,
    2009EPJD...55..127Z, 2012A&A...537A.124Z}.
Its derivation starts with the ansatz that any perturbation $\delta f(r, \theta, z; t)$ to the equilibrium $f(r)$ takes the form
\begin{eqnarray}
 \delta f(r, \theta, z; t) = \mathrm{Re}\left\{\tilde{f}(r)\exp\left[{\rm i}\left(kz + n\theta -\omega t\right)\right]\right\} ,
\label{eq_wav_ansatz}
\end{eqnarray}
  where $\mathrm{Re}(\cdots)$ means taking the real part of the function.
Besides, $k$ and $n$ are the longitudinal and azimuthal wavenumbers, respectively.
The phase speed $\vph$ is defined as $\vph=\omega/k$.
One proceeds by defining
\begin{eqnarray}
 m_i^2 = k^2 \frac{\left[c_i^2 - \left(\vph -U_i\right)^2\right]\left[v_{{\rm A}i}^2 - \left(\vph -U_i\right)^2\right]}
   {\left(c_i^2 + v_{{\rm A} i}^2\right) \left[c_{{\rm T} i}^2 - \left(\vph -U_i \right)^2\right]} ,
\label{eq_def_m2}
\end{eqnarray}
where $i=0, {\rm e}$.
To ensure the waves are trapped, $m_{\rm e}^2$ has to be positive, meaning that
\begin{eqnarray}
 c_{\mathrm{m},{\rm e}}^2 < (\vph-U_{\rm e})^2 < c_{\mathrm{M},{\rm e}}^2 \text{ or }
    (\vph-U_{\rm e})^2 < c_{{\rm Te}}^2 ,
\end{eqnarray}
where $c_{\mathrm{m},{\rm e}} = \mathrm{min}(c_{\rm e}, v_{\rm Ae})$
 and $c_{\mathrm{M},{\rm e}} = \mathrm{max}(c_{\rm e}, v_{\rm Ae})$.
On the other hand, the spatial profiles of the perturbations
    in the $r-$ direction are determined by the sign of $m_0^2$.
When $m_0^2 <0$ ($m_0^2 >0$), the waves are body (surface) ones,
    corresponding to an oscillatory (a spatially decaying) $r$- dependence
    inside the cylinder.

With Eq.(\ref{eq_wav_ansatz}) inserted into the linearized, ideal MHD equations, the DR follows
    from the requirements that the radial component of the Lagrangian displacement $\xi_r$
    and the total pressure $\delta p_{\rm T}$ be continuous at $r=a$.
The DR reads
\begin{eqnarray}
 \frac{\rho_{\rm e}}{\rho_0}\frac{m_0}{|m_{\rm e}|}
   \frac{\left[v_{\rm Ae}^2 - \left(\vph - U_{\rm e}\right)^2\right]}{\left[v_{{\rm A}0}^2 - \left(\vph - U_0\right)^2\right]}
   \frac{I_n'(m_0 a)}{I_n(m_0 a)}
 = \frac{K_n'(|m_{\rm e}| a)}{K_n(|m_{\rm e}| a)}
\label{eq_DR_surface}
\end{eqnarray}
for surface waves,
and
\begin{eqnarray}
 \frac{\rho_{\rm e}}{\rho_0}\frac{n_0}{|m_{\rm e}|}
   \frac{\left[v_{\rm Ae}^2 - \left(\vph - U_{\rm e}\right)^2\right]}{\left[v_{{\rm A}0}^2 - \left(\vph - U_0\right)^2\right]}
   \frac{J_n'(n_0 a)}{J_n(n_0 a)}
 = \frac{K_n'(|m_{\rm e}| a)}{K_n(|m_{\rm e}| a)}
\label{eq_DR_body}
\end{eqnarray}
for body waves ($n_0^2 = -m_0^2 > 0$).
Furthermore, kink and sausage waves correspond to the solutions to the DR with $n$ being $1$ and $0$, respectively.
The prime denotes the derivative of Bessel function with respect to its argument,
   e.g., $J_n'(n_0 a) \equiv {\rm d} J_n(x)/{\rm d} x$ with $x = n_0 a$.
One may note that $m_{\rm e}$ appears only as absolute values to ensure that the waves external to the cylinder are evanescent.

It proves necessary to examine
    the importance of  density fluctuation  relative to
    the transverse displacement.
This is readily done by evaluating $X \equiv \left.(\tilde{\rho}/\rho_0 )/(\tilde{\xi}_r/a)\right|_{r=a}$,
\begin{eqnarray}
X =\left. \frac{ (m_0^2 a) (\omega - k U_0)^2}{\left[k^2 c_0^2 - (\omega - k U_0)^2\right]} \frac{\tilde{p}_{\rm T}}{{\rm d}\tilde{p}_{\rm T}/{\rm d}r} \right |_{r=a} .
\label{eq_ratio_rho_xi}
 \end{eqnarray}
For body waves, $\tilde{p}_{\rm T}$ inside the cylinder is proportional to $J_1(n_0 r)$ for a kink wave, and to
    $J_0(n_0 r)$ for a sausage one, resulting in
\begin{eqnarray}
X=
 \frac{(\vph - U_0)^2}{(\vph - U_0)^2 - c_0^2}
 \left\{\begin{array}{l}
  (n_0 a) J_1(n_0 a)/J_1'(n_0 a) \\
		\hskip 0.5cm \text{kink},  \\
  (n_0 a) J_0(n_0 a)/J_0'(n_0 a) \\
		\hskip 0.5cm  \text{sausage}.
		\end{array} \right.
\label{eq_Bdy_rhoOxi}
\end{eqnarray}
Likewise, for surface waves, inside the cylinder $\tilde{p}_{\rm T} \propto I_1(m_0 r)$ for a kink wave,
    and $\propto I_0(m_0 r)$ for a sausage one, leading to
\begin{eqnarray}
X =
  \frac{(\vph - U_0)^2}{c_0^2 - (\vph - U_0)^2}
  \left\{\begin{array}{l}
  (m_0 a) I_1(m_0 a)/I_1'(m_0 a)  \\
		\hskip 0.5cm \text{kink},  \\
  (m_0 a) I_0(m_0 a)/I_0'(m_0 a) \\
		\hskip 0.5cm  \text{sausage}.
		\end{array} \right.
\label{eq_Srfc_rho0xi}
\end{eqnarray}

The dispersion relations~(\ref{eq_DR_surface}) and (\ref{eq_DR_body})
   possess three symmetric properties that allow us to simplify our examination of the standing modes.
The first two dictate that
   if $[\vph, k; U_0, U_{\rm e}]$ represents a solution to the DR, then so does $[\vph, -k; U_0, U_{\rm e}]$;
   if $[\vph, k; U_0, 0]$ is a solution, then so is $[-\vph, k; -U_0, 0]$ (see Eq.(\ref{eq_def_m2}) with $U_{\rm e} = 0$).
They were detailed in the appendix of paper I which adopts a slab geometry,
   and can be readily shown to hold in the cylindrical case if one recognizes that
   $x Z_n'(x)/Z_n(x)$ is an even function for Bessel functions $Z_n$ of integer order $n$,
   where $Z_n$ is $J_n$ or $I_n$.
They are summarized here for one to realize that as long as the external medium is at rest ($U_{\rm e}=0$),
   as will be assumed throughout this study,
   then for the purpose of examining how the period ratio depends on the internal flow $U_0$,
   one needs only to consider positive $U_0$.
The third symmetry property reflects simply a Galilean transformation, which relates the phase speed
    $\vph(k; U_0, U_{\rm e})$ in one frame, where the speeds read $U_0$ and $U_{\rm e}$,
    to $\vph(k; U_0^\dagger, U_{\rm e}^\dagger)$ in a different one where the speeds read $U_0^\dagger$ and $U_{\rm e}^\dagger$.
Certainly one requires that $U_{\rm e}^\dagger-U_0^\dagger = U_{\rm e} - U_0$.
One then sees that $\vph(k; U_0^\dagger, U_{\rm e}^\dagger) = \vph(k; U_0, U_{\rm e})+(U_0^\dagger-U_0)$,
    and in particular, $\vph(k; U_0-U_{\rm e}, 0) = \vph(k; U_0, U_{\rm e})-U_{\rm e}$.
What this means is that, even though the wave dispersion properties
   expressed as a series of analytical expressions in a number of physically interesting limits
   in both coronal and photospheric environments are to be derived in a frame where $U_{\rm e}=0$,
   they can be easily extended to an arbitrary frame of reference.

\section{Period ratios for standing modes supported by coronal cylinders}
\label{sec_Cor_Pratio}

\subsection{Overview of Coronal Cylinder Dispersion Diagrams}
\label{sec_sub_Cor_DR}

Consider first the coronal case, where the ordering $v_{\rm Ae}>v_{{\rm A}0}>c_0>c_{\rm e}$ holds.
To be specific, we choose $v_{{\rm A}0}=4 c_0$ and $c_{\rm e}=0.72 c_0$, the observational justification of which was given in paper I.
For the external Alfv\'en speed, unless otherwise specified, we will discuss in detail
   a reference case where $v_{\rm Ae} = 2 v_{{\rm A}0}$.
Evidently, the larger the ratio $v_{\rm Ae}/v_{{\rm A}0}$, the stronger the density contrast.
\begin{figure}
\centerline{\includegraphics[width=0.6\columnwidth]{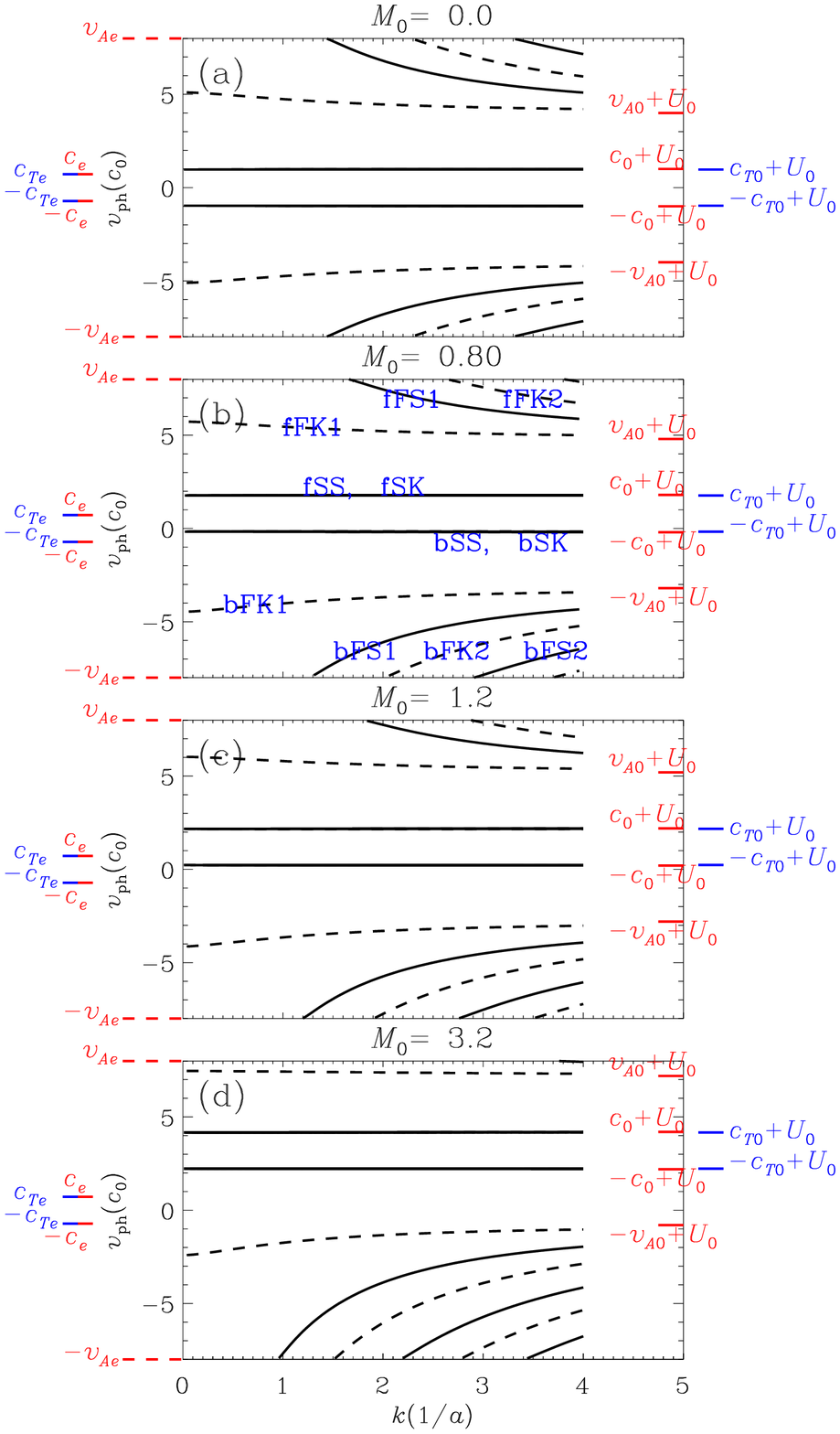}}
\caption{Phase speeds $\vph$ as a function of longitudinal wavenumber $k$ for a series of internal flow $U_0$.
Expressing $U_0$ in units of the internal sound speed $U_0 = M_0 c_0$,
   panels (a) to (d) correspond to an $M_0$ of $0, 0.8, 1.2$ and $3.2$, respectively.
On the left (right) of each panel, the characteristic speeds external (interior) to the cylinder are given
   by the horizontal bars.
In particular, the external Alfv\'en speed provides the lower and upper bounds, as indicated by the long red dashed bars.
Kink and sausage modes are presented by the dashed and solid curves, respectively.
They are further labeled, as shown in panel (b), using combinations of letters b/f+F/S+K/S, representing
   backward or forward, Fast or Slow, Kink or Sausage.
The number appended to the letters denote the order of occurrence.
Hence, bFK1 represents the first branch of backward Fast Kink mode.
Moreover, here $v_{\rm Ae}=8 c_0$, $c_{\rm e}=0.72c_0$, $c_{{\rm Te}}=0.719c_0$, while $v_{{\rm A}0}=4c_0$, and $c_{{\rm T}0}=0.97c_0$.
}
 \label{fig_Cor_DR}
\end{figure}

Figure~\ref{fig_Cor_DR} presents the dependence on longitudinal wavenumber $k$ of the phase speeds $\vph$
    for a series of $U_0 = M_0 c_0$, where the internal Mach number $M_0$ reads $0, 0.8, 1.2$
    and $3.2$, respectively.
Kink and sausage waves are plotted with the dashed and solid curves, respectively.
As shown in Fig.\ref{fig_Cor_DR}b, they are labeled by combinations of letters b/f+F/S+K/S, representing
    backward or forward, Fast or Slow, Kink or Sausage.
``Fast'' or ``Slow'' is related to the magnitude of the phase speed,
    while ``backward'' or ``forward'' derives from the sign of the phase speeds when the flow is absent,
    and was termed ``originally backward-(forward-) propagating" by~\citet{2000ApJ...531..561A} in the same sense.
The number appended to the letters denotes the order of occurrence, meaning that
    fFK1 represents the first branch of forward Fast Kink wave.
The characteristic speeds external (interior) to  the cylinder are given on the left (right) of each panel
    to aid wave categorization.
In agreement with~\citet{2003SoPh..217..199T}(hereafter TEB03), Fig.\ref{fig_Cor_DR} indicates that
    all waves in such a coronal environment are body waves.

A clear flow dependence can be seen in Fig.\ref{fig_Cor_DR}.
Consider first the slow waves.
The propagation windows always encompass $(-c_{0}+U_0, -c_{{\rm T}0}+U_0)$ and $(c_{{\rm T}0}+U_0, c_{0}+U_0)$,
    which is readily understandable when one examines the slender and thick cylinder limits.
In the former limit ($ka \ll 1$),
\begin{eqnarray}
 {\vph} \approx U_0 \pm c_{{\rm T}0} \sqrt{1+ \frac{c_{{\rm T}0}^4}{c_0^2 v_{{\rm A}0}^2} \frac{k^2 a^2}{h_{l,\pm}^2}} ,
 \label{eq_Cor_slow_smallk}
\end{eqnarray}
where $h_{l,\pm}$ has an infinite number of values.
For kink waves, $h_{l,\pm}$ is an arbitrary root of the transcendental equation
\begin{eqnarray}
\frac{x J_1'(x)}{J_1 (x)} = -\frac{\rho_0}{\rho_{\rm e}}\frac{v_{{\rm A}0}^2-c_{{\rm T}0}^2}{v_{\rm Ae}^2 - c_{{\rm T}0,\pm}^2} ,
\label{eq_hj_KINK}
\end{eqnarray}
   where $c_{{\rm T}0,\pm} = \pm c_{{\rm T}0}+U_0$, and $x$ denotes the unknown.
For sausage waves, $h_{l,\pm}$ can be approximated by
\begin{eqnarray}
h_{l,\pm} \approx j_{1,l} .
\label{eq_hj_sausage}
\end{eqnarray}
When the opposite limit holds ($ka \gg 1$), one finds
\begin{eqnarray}
 \vph \approx  U_0 \pm c_{0} \sqrt{1- \frac{c_{0}^2}{v_{{\rm A}0}^2 - c_0^2} \frac{g_l^2}{k^2 a^2}},
 \label{eq_Cor_slow_bigk}
\end{eqnarray}
where
\begin{eqnarray}
   g_l =\left\{\begin{array}{l c}
           j_{1,l} 	  & \mbox{      kink} \\
           j_{0,l} & \mbox{      sausage}
          \end{array} \right.
\label{eq_gj}
\end{eqnarray}
in which $l=0, 1, 2,\cdots$, and $j_{n, l}$ denotes the $l$-th zero of $J_n$.
The plus and minus signs in Eqs.(\ref{eq_Cor_slow_smallk}) and (\ref{eq_Cor_slow_bigk}) correspond to
   the upper and lower bands, respectively.
However, the slow waves in the coronal case are not of interest as far as the period ratio $P_1/2P_2$ is concerned,
   for they are nearly dispersionless due to the nearly indistinguishable values of $c_0$ and $c_{{\rm T}0}$,
   and the deviation of $P_1/2P_2$ from unity in the present study derives entirely from wave dispersion.

In view of their stronger dispersion, let us pay a closer look at fast waves whose propagation windows
   encompass $(-v_{\rm Ae}, U_0-v_{{\rm A}0})$ and $(U_0+v_{{\rm A}0}, v_{\rm Ae})$.
One may readily understand this by examining the thick cylinder limit ($ka \gg 1$), where one finds
\begin{eqnarray}
 \vph \approx  U_0 \pm v_{{\rm A}0} \sqrt{1+ \frac{v_{{\rm A}0}^2}{v_{{\rm A}0}^2 - c_0^2} \frac{h_{l,\pm}^2}{k^2 a^2}}.
 \label{eq_Cor_fast_bigk}
\end{eqnarray}
With the exception of bFK1 and fFK1, there exist wavenumber cutoffs for both kink and sausage waves, and these are given by
\begin{eqnarray}
 (ka)_{\rm c} = g_l \Lambda_{\pm},
\label{eq_Cor_fast_cutoff}
\end{eqnarray}
where
\begin{eqnarray*}
 \Lambda_{\pm} = \sqrt{\frac{(c_0^2+v_{{\rm A}0}^2)[(v_{\rm Ae}\mp U_0)^2-c_{{\rm T}0}^2]}{[(v_{\rm Ae}\mp U_0)^2-c_{0}^2][(v_{\rm Ae}\mp U_0)^2-v_{{\rm A}0}^2]}}  .
\end{eqnarray*}
On the other hand, for bFK1 and fFK1 in the slender cylinder limit $ka \ll 1$, $\vph$ may be approximated by
\begin{eqnarray}
 \vph^{\pm} \approx  d_\pm
 \left\{1 \pm \frac{\hat{\rho}_0 \left[ v_{{\rm A}0}^2 - \left(d_\pm - U_0\right)^2\right]}{2 d_\pm d_{k}}
    (\lambda_{\pm} k a)^2 K_0 (\lambda_{\pm} |k|a)
 \right\},
 \label{eq_Cor_FstKnk_smallK}
\end{eqnarray}
    where
\begin{subequations}
\label{eq_def_d_lambda}
\begin{align}
& d_{\pm} = \hat{\rho}_0 U_0 \pm d_k, \\
& d_k     = \sqrt{c_k^2 - \hat{\rho}_0 \hat{\rho}_{\rm e} U_0^2},  \label{eq_dk} \\
& \lambda_{\pm} = \sqrt{\frac{(d_{\pm}^2-c_{\rm e}^2)(v_{\rm Ae}^2 - d_{\pm}^2)}{(c_{\rm e}^2+v_{\rm Ae}^2)(d_{\pm}^2-c_{{\rm Te}}^2)}} .
\end{align}
\end{subequations}
Moreover, $\vph^{+}$ and $\vph^{-}$ represent the upper and lower branches, respectively.

Compared with available ones,
   our study offers some new analytical expressions for the phase speed $\vph$ in a number of physically interesting limits.
Equations~(\ref{eq_Cor_slow_smallk}) and (\ref{eq_Cor_slow_bigk}) offer the approximate expressions of $\vph$
   for slow waves in the slender and thick cylinder limits, respectively.
For the fast ones, Eq.(\ref{eq_Cor_fast_bigk}) presents an explicit expression for $\vph$ in the limit of $ka\gg 1$,
   thereby extending the original discussion of static cylinders in this situation by~\citet{1983SoPh...88..179E} (hereafter ER83)
   where the authors emphasized the analogy with the Love waves of seismology and Pekeris waves of oceanography (see Eq.(13) in ER83).
Moreover, Eq.(\ref{eq_Cor_FstKnk_smallK}) examines fast kink waves in the slender cylinder limit $ka \ll 1$,
   and extends available results in three ways.
First, neglecting the first order correction, Eq.(\ref{eq_Cor_FstKnk_smallK}) reduces to $d_\pm$, which agrees
   with Eq.(70) in~\citet{1992SoPh..138..233G}.
Second, taking $U_0 =0$, we recover the expression for a static cylinder, namely Eq.(15) in ER83.
Our expression also shows that Eq.(15) as given in ER83 is in fact not restricted to the cold plasma limit ($c_{\rm e} = c_0 = 0$),
   but valid for a rather general coronal environment as long as $\lambda$ is generalized to incorporate $c_{\rm e}$ and $c_{\rm Te}$,
   as given by our Eq.(\ref{eq_def_d_lambda}).
Third, the plus version $\vph^{+}$ reduces to Eq.(5) in~\citet{2009A&A...498L..29V} where the transverse waves propagating in soft X-ray coronal jets are examined,
   when one notes that $(d_{+}^2-c_{\rm e}^2)/(d_{+}^2-c_{{\rm Te}}^2) \approx 1$, and $v_{\rm Ae}^2 \gg c_{\rm e}^2$.
However, it turns out that except for extremely small $ka$, retaining the original form in terms of the modified Bessel function $K_0$
   is more accurate than the logarithmic form given by Eq.(5) in~\citet{2009A&A...498L..29V}.
Furthermore, the expression $\vph^{-}$ gives the phase speed for the waves that are backward propagating in the absence of flow.

\subsection{Procedures for Computing Standing Modes}
\label{sec_sub_method}

By ``standing'', we require that the radial Lagrangian displacement $\xi_r (r, \theta, z; t)$ is zero
    at the interface $r=a$ at both ends of the cylinder $z=0, L$, irrespective of $\theta$ and $t$.
One requirement for this to be true for arbitrary $\theta$ is that only propagating waves with identical azimuthal wavenumbers $n$
    can combine to form standing modes.
A pair of propagating waves characterized by a common angular frequency $\omega$ but different
    longitudinal wavenumbers $k_r$ and $k_l$ then lead to that
\begin{eqnarray}
\xi_r(r, \theta, z; t) =
  &&  \mathrm{Re}\left\{\tilde{\xi}_{r,l}(r)\exp\left[{\rm i}\left(k_l z + n\theta -\omega t \right)\right] \right\} \nonumber \\
  &+& \mathrm{Re}\left\{\tilde{\xi}_{r,r}(r)\exp\left[{\rm i}\left(k_r z + n\theta -\omega t \right)\right] \right\} .
\label{eq_displacement}
\end{eqnarray}
Specializing to $(r, z)=(a, 0)$, one finds $\tilde{\xi}_{r,l}(a) = - \tilde{\xi}_{r,r}(a)$, meaning that
   one is allowed to choose $\tilde{\xi}_{r,l}(a) = A_{\xi}$ to be real.
It then follows that
\begin{eqnarray}
&&  \xi_r(a, \theta, z; t) \nonumber \\
&=& A_\xi \left[\cos\left(\omega t - k_l z - n\theta \right) - \cos\left(\omega t - k_r z - n\theta\right)\right] \nonumber \\
&=& -2A_\xi\sin\left(\frac{k_r-k_l}{2}z\right)\sin\left(\omega t-\frac{k_l + k_r}{2}z - n\theta\right) .
\label{eq_stand_displc}
\end{eqnarray}
For $\xi_r(a, \theta, L; t)$ to be zero at arbitrary $t$, this requires
\begin{eqnarray}
k_r - k_l =  \frac{2\pi m}{L}, m=1, 2, \cdots
\label{eq_k4stand}
\end{eqnarray}
By convention, $m=1$ corresponds to the fundamental mode, and $m=2$ to its first overtone.

At this point, it suffices to say that the procedure for computing the period ratios of standing modes
   is identical to the slab case, which was detailed in paper I.
Basically it involves constructing an $\omega-k$ diagram where each propagating wave in a pair to form
   standing modes corresponds to a particular curve, meaning that
   a horizontal cut with a constant $\omega$ would intersect with the two resulting curves at two points.
If the separation between the two points is $2\pi/L$, then one finds the fundamental mode.
If it is twice that, one finds the first overtone.
Let the angular frequency of the fundamental mode (first overtone) be denoted by $\omega_1$ ($\omega_2$),
   the period ratio is simply $P_1 /2 P_2 = \omega_2 /2 \omega_1$.
The existence of cutoff wavenumbers for sausage waves to be trapped
   translates into the existence of cutoff
   aspect ratios $(a/L)_{\mathrm{cutoff}}$ for standing sausage modes to be non-leaky.
As emphasized by paper I (see Fig.3 therein), this $(a/L)_{\mathrm{cutoff}}$ is
   not determined by the difference between the two
   cutoffs of the two $\omega - k$ curves divided by $2\pi$, but larger than that.

When computing the coronal standing modes, we consider only bFK1 and fFK1 for kink modes,
   and bFS1 and fFS1 for sausage modes.
Branches with larger mode numbers like bFK2 or bFS2 would form standing modes only
   for relatively thick cylinders where $a/L$ is of the order unity.
For the same reason, we discard the combinations between slow and fast sausage propagating waves.
On the other hand, combinations of slow with fast kink wave, such as bFK1 plus fSK,
   turn out extremely unlikely as well.
This is because, while slow kink waves are dominated by the intensity oscillations instead of transverse displacements ($|X|\gg 1$),
   the opposite holds for fast ones ($|X|\ll 1$).
The end result is that if a fast kink wave does combine with a slow one to form a standing mode,
   a transverse loop displacement on the order of the cylinder radius will lead to a relative intensity variation
   that exceeds unity.
To see this, consider slender cylinders such that $k a \rightarrow 0$, and consider the case where the components to form standing modes
   are bFK1 and any branch of fSK.
For bFK1, one sees that $\vph \approx d_{-}$ and $n_0 a \rightarrow 0$, and hence
    $X \approx \frac{(d_{-} - U_0)^2}{(d_{-} - U_0)^2 - c_0^2}(n_0 a)^2$.
Because $d_{-} - U_0 \approx -d_{k}-\hat{\rho}_{\rm e} U_0$ is of the order of $v_{{\rm A}0}$,
    and $v_{{\rm A}0}^2 \gg c_0^2$,
    $X$ would be roughly $(n_0 a)^2$ and hence approaches zero as well.
However, for slow kink waves, by noting that $(\vph - U_0)^2 \rightarrow c_{{\rm T}0}^2$ when $ka\rightarrow 0$,
    one finds that
    $X \approx (n_0 a)^2 (1-\vph^2/v_{\rm Ae}^2)(\rho_{\rm e} v_{\rm Ae}^2)/(\rho_0 c_{{\rm T}0}^2)$,
    which is approximately $(n_0 a)^2 v_{{\rm A}0}^2/c_0^2$ since $\rho_{\rm e} v_{\rm Ae}^2 \approx \rho_0 v_{{\rm A}0}^2$ and $c_{{\rm T}0}\approx c_0$.
Note that in coronal conditions, $h_{l, \pm}$ as given by Eq.(\ref{eq_hj_KINK}) can be approximated by $(l+3/4)\pi$.
When $ka \rightarrow 0$, with $n_0 a$ approaching $h_{l, \pm}$, $X$ will be large.

\subsection{Period Ratios for Standing Kink Modes}
\label{sec_sub_Cor_PrKnk}

\begin{figure}
\centerline{\includegraphics[width=0.6\columnwidth]{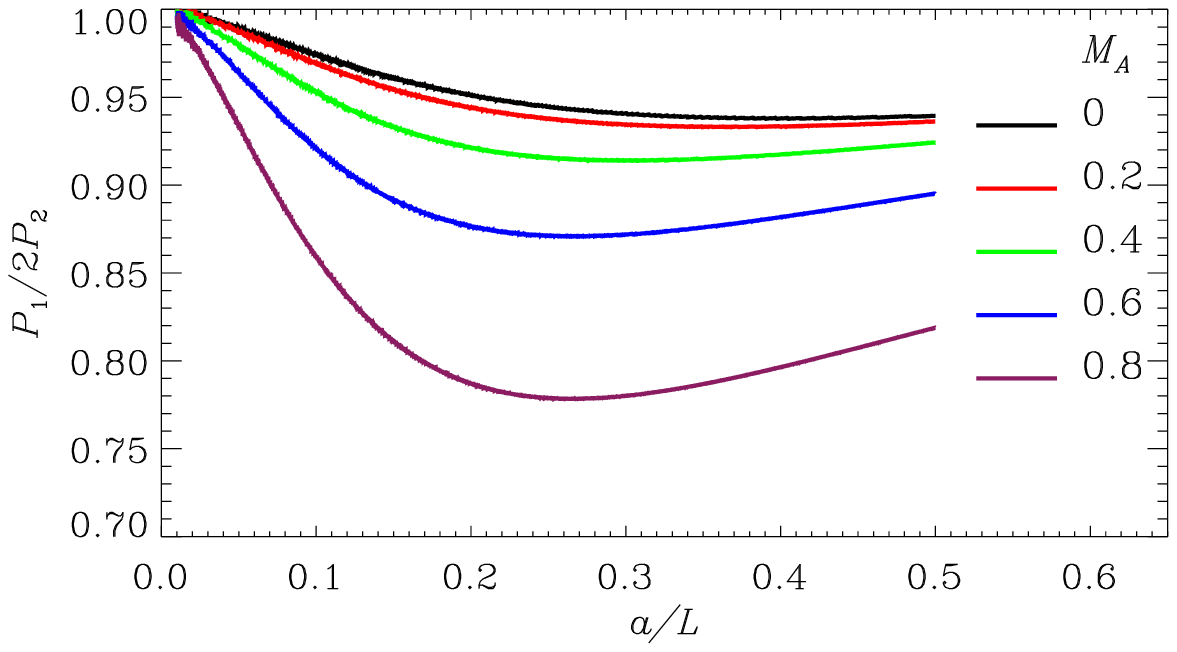}}
 \caption{Period ratio $P_1/2P_2$ as a function of the cylinder aspect
   ratio $a/L$ for standing fast kink modes.
Curves with different colors represent results computed for different values of the
   flow speed $U_0$ measured in units of the internal Alfv\'{e}n speed
   $U_0 = M_{\rm A} v_{{\rm A}0}$.
}
 \label{fig_Cor_PrKnk}
\end{figure}

Figure~\ref{fig_Cor_PrKnk} presents the dependence on the aspect ratio $a/L$ of the period ratio $P_1/2P_2$
   pertinent to standing fast kink modes.
Here the results for a number of different $U_0$ are shown with different colors, with $U_0$ represented by the internal
   Alfv\'en Mach number $M_{\rm A} = U_0/v_{{\rm A}0}$.
One can see that all curves decrease from unity at zero $a/L$, attain a minimum, and then increase towards unity.
Increasing $U_0$ substantially strengthens the deviation of $P_1/2P_2$ from unity relative to the static case
   (the black curve).
Take the minima for instance.
While in the static case it reads $0.938$, attained at $a/L=0.405$, when $M_{\rm A}=0.8$ it is significantly reduced
   to $0.778$ attained at $a/L=0.267$.
At smaller aspect ratios, the dispersion introduced by the flow, and hence the deviation from unity of
   the period ratio $P_1/2P_2$, is not as strong.
However, at an aspect ratio of $a/L = 0.19$, one finds that $P_1/2P_2$ decreases significantly from
   $0.953$ in the static case to $0.79$ when $M_{\rm A} = 0.8$.
Actually this aspect ratio corresponds to the $NoRH$ loop that experienced standing
   kink oscillations on 2002 July 3 with multiple periodicities  that yield $P_1/2P_2 = 0.82$~\citep{2013SoPh..284..559K}.
One finds that while the wave dispersion due to transverse density structuring alone cannot account
   for this measured value of $P_1/2P_2$, it may be attained with the aid of the additional wave dispersion due to flow shear.
In this regard, we agree with~\citet{2009SSRv..149....3A} in the sense that the contribution of the density contrast
   alone to the deviation of $P_1/2P_2$ from unity seems to be marginal for extremely thin cylinders.
However, we note that when a substantial flow shear exists between the cylinder and its surroundings,
    the shear-associated wave dispersion may not be neglected for loops with finite aspect ratios.
As a matter of fact, for loops with $a/L$ as small as $0.05$, the flow effect is still substantial enough to
    be of observational significance: while $P_1/2P_2$ reads $0.989$ in the static case, when $M_{\rm A}=0.8$ it is $0.934$, which is
   already below the minimum $P_1/2P_2$ can reach when the flow is absent.
We note that this $a/L$ is not unrealistic but lies within the range of the measured values
    of oscillating EUV loops examined in~\citet{2002ApJ...576L.153O}(see their Table 1).
The point we want to make here is that the wave dispersion associated with the transverse structuring
    needs to be considered for a theoretical understanding of the period ratios of standing kink modes,
    and this is particularly necessary in the presence of a strong flow shear and
    when the loop aspect ratios are not extremely small.

\begin{figure}
\centerline{\includegraphics[width=0.6\columnwidth]{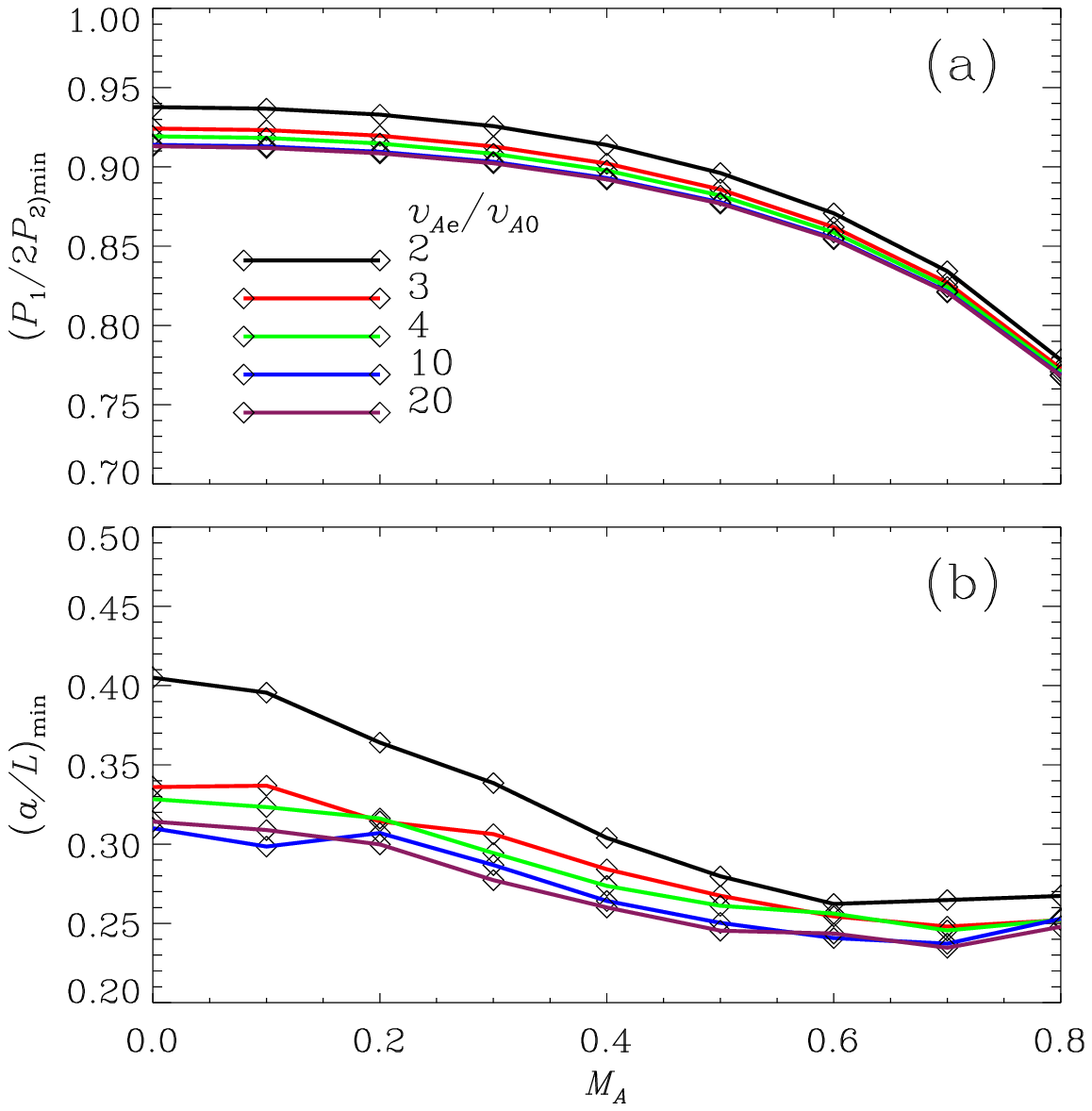}}
 \caption{Effects of flow speed on (a) the minimal period ratio, $(P_1/2P_2)_{\mathrm{min}}$, and (b) the aspect ratio
    at which the minimum is attained, $(a/L)_{\mathrm{min}}$.
Here both $(P_1/2P_2)_{\mathrm{min}}$ and $(a/L)_{\mathrm{min}}$
    are displayed as a function of the Alfv\'enic Mach number $M_{\rm A}$.
}
 \label{fig_Cor_PrKnk_MA}
\end{figure}
Figure~\ref{fig_Cor_PrKnk_MA} further examines the flow effect by showing
    (a) the minimal period ratio, $(P_1/2P_2)_{\mathrm{min}}$ and (b) its location, $(a/L)_{\mathrm{min}}$,
    as a function of the internal Alfv\'enic Mach number $M_{\rm A}$.
In addition to the reference case where $v_{\rm Ae}/v_{{\rm A}0}=2$, Fig.\ref{fig_Cor_PrKnk_MA} also examines
    other ratios of $3$, $4$, $10$, and $20$, shown in different colors.
Regarding Fig.\ref{fig_Cor_PrKnk_MA}a, one sees that the flow effect on
    the period ratios is significant for all the considered $v_{\rm Ae}/v_{{\rm A}0}$, or equivalently, the density contrast.
As a matter of fact, at a given $M_{\rm A}$, even though $(P_1/2P_2)_{\mathrm{min}}$ tends to decrease with
    increasing density contrast,
    this tendency is rather weak, and seems to saturate when $v_{\rm Ae}/v_{{\rm A}0}$ exceeds, say, 10,
    as evidenced by the fact that the two curves corresponding to $v_{\rm Ae}/v_{{\rm A}0}$ being $10$ and $20$
    can hardly be distinguished.
Consequently, when $v_{\rm Ae}/v_{{\rm A}0}$ is as large as $20$, $(P_1/2P_2)_{\mathrm{min}}$
    decreases from $0.914$ in the static case to $0.768$ when $M_{\rm A}=0.8$, amounting to
    a relative difference of $16\%$, which is almost the same as in the case when $v_{\rm Ae}/v_{{\rm A}0}$ is $2$
    where this fractional difference reads $17.1\%$.
Looking at Fig.\ref{fig_Cor_PrKnk_MA}b, one notices that for a given density contrast, the aspect ratio at which
    the minimum period ratio is attained tends to decrease with increasing flow,
    and this tendency is clearer for weaker density contrasts.
When $v_{\rm Ae}/v_{{\rm A}0}$ is at the two extremes, $(a/L)_{\mathrm{min}}$
    reads $0.405$ and $0.31$ in the static case,
    and goes down to $0.267$ and $0.248$ for an $M_{\rm A}$ being $0.8$, respectively.
The fractional change due to the flow in the former reads 34\%,
    while 20\% in the latter.

It is interesting to contrast the cylinder case with the slab one.
In both cases the minimal period ratio $(P_1/2P_2)_{\mathrm{min}}$ and
    the aspect ratio $(a/L)_{\mathrm{min}}$ have been examined analytically.
Note that in the slab case, $a$ refers to the half-width of the slab.
For cold static slabs, \citet{2011A&A...526A..75M} established that $(P_1/2P_2)_{\mathrm{min}}$ can never drop below $\sqrt{2}/2$, which is
    attained for the infinite density contrast at a zero aspect ratio.
While this was established by employing the Epstein profile to connect the slab density and the density of its surroundings,
    the numerical results in both~\citet{{2011A&A...526A..75M}} and paper I demonstrate that this lower limit for $P_1/2P_2$
    is also valid when the density profile is in the form of a step function.
When a flow $U_0$ is introduced, paper I shows that $P_1/2P_2$ is no longer subject to this lower limit
    and the change in $P_1/2P_2$ relative to the static case is typically $\sim 20\%$.
Besides, $(a/L)_{\mathrm{min}}$ tends to increase with increasing $U_0$.
For cold static cylinders, \citet{2006A&A...460..893M} (hereafter M06) and also \citet{2009SSRv..149....3A}
    established that $(P_1/2P_2)_{\mathrm{min}}$ also suffers from
    a lower limit of $\sim 0.92$ when the density contrast approaches infinity,
    and the aspect ratio where this lower limit is attained is $\sim 0.3$
    (see Figure 2 in M06, and note that the symbol $L$ therein is the loop half-length, and hence
    their $a/L$ corresponds to twice the value of $a/L$ in the present study).
The static case in Fig.\ref{fig_Cor_PrKnk} agrees remarkably well with Fig.2 in M06 despite that
    the sound speeds are allowed to be non-zero
    now, which is not surprising given that the sound speeds are significantly smaller than the Alfv\'en speeds.
However, Fig.\ref{fig_Cor_PrKnk} offers the new result that in the cylinder case,
    the introduction of the internal flow provides significant revision to the period ratio,
    making it no longer suffer from the lower limit established for static cylinders.
This is true even when the density contrast approaches infinity,
    and the revision to the period ratio is typically $\sim 16-17\%$, similar to the slab case.
At a given $v_{\rm Ae}/v_{{\rm A}0}$, the tendency for $(a/L)_{\mathrm{min}}$ to decrease with increasing $U_0$
    in the cylinder case is opposite to what happens for slabs with flows.

\subsection{Period Ratios for Standing Sausage Modes}
\label{sec_sub_pratio_sausage}

\begin{figure}
\centerline{\includegraphics[width=0.6\columnwidth]{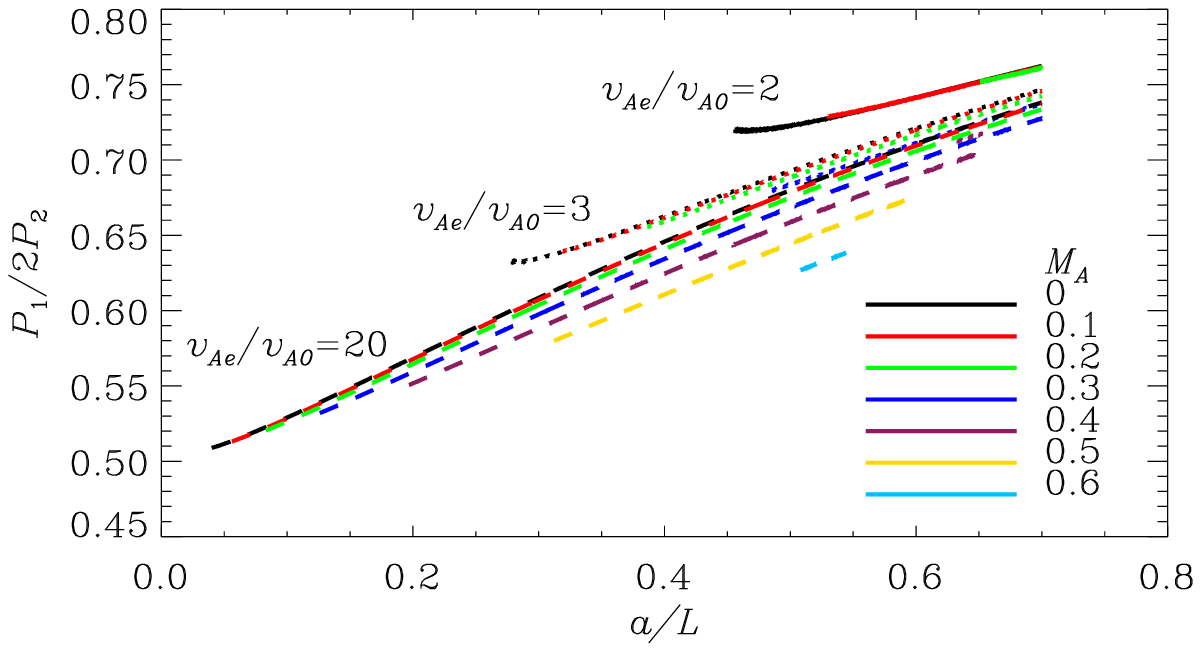}}
 \caption{Period ratio $P_1/2P_2$ as a function of the cylinder aspect
   ratio $a/L$ for standing sausage modes.
The solid, dotted, and dashed curves are for the cases where $v_{\rm Ae}/v_{{\rm A}0}=2$, $3$, and $20$, respectively.
Curves with different colors represent results computed for different values of the
   flow speed $U_0$ measured in units of the internal Alfv\'en speed
   $U_0 = M_{\rm A} v_{{\rm A}0}$.
}
 \label{fig_Cor_PrSsg}
 \end{figure}

Figure~\ref{fig_Cor_PrSsg} presents the period ratio $P_1/2P_2$ as a function of aspect ratio $a/L$
   for a series of $v_{\rm Ae}/v_{{\rm A}0}$, pertinent to standing sausage modes.
The solid, dotted, and dashed curves correspond to $v_{\rm Ae}/v_{{\rm A}0}$ being $2, 3$,
   and $20$, respectively.
As indicated by the different colors, a set of $U_0$ is investigated
   and measured in units of the internal Alfv\'en speed $U_0 = M_{\rm A} v_{{\rm A}0}$.
It is clear from Fig.\ref{fig_Cor_PrSsg} that the effect of flow on the period ratio $P_1/2P_2$
   is not as strong as for the kink modes.
Since this effect increases with increasing $v_{\rm Ae}/v_{{\rm A}0}$, one may examine the extreme where $v_{\rm Ae}/v_{{\rm A}0} = 20$,
   in which case one finds that at $a/L=0.4$, $P_1/2P_2$ reads $0.611$ when $M_{\rm A}=0.5$, which is 5.4\% lower than the value $0.646$ obtained in the static case.
This fractional change in $P_1/2P_2$ is typical in this case at a given aspect ratio.
However, the flow effect is much stronger when it comes to the cutoff aspect ratio $(a/L)_{\mathrm{cutoff}}$
   only above which can standing sausage modes be supported.
This effect is substantial even when it is the weakest among the three $v_{\rm Ae}/v_{{\rm A}0}$ considered:
   when $v_{\rm Ae}/v_{{\rm A}0}=2$, $(a/L)_{\mathrm{cutoff}}$ increases from $0.456$ to $0.53$ to $0.651$
   with $M_{\rm A}$ increasing from $0$ to $0.1$ to $0.2$.
Regarding the other extreme $v_{\rm Ae}/v_{{\rm A}0}=20$,
   while $(a/L)_{\mathrm{cutoff}}$ reads $0.04$ for the static case, it reads $0.083$ when $M_{\rm A}=0.2$ and $0.197$ when $M_{\rm A}=0.4$.
This means that at a given $v_{\rm Ae}/v_{{\rm A}0}$, relative to the static case, cylinders with flow can support
   standing sausage modes only when they are sufficiently thicker if the cylinder length is fixed.

At this point, a comparison with studies of sausage modes supported by magnetized slabs is informative.
As demonstrated numerically by~\citet{2009A&A...503..569I} and analytically by~\citet{2011A&A...526A..75M},
    for static coronal slabs $P_1/2P_2$ may reach as low as $1/2$ with
    the lower limit attainable when the density contrast is infinite.
Besides, the cutoff aspect ratio lowers with increasing density contrast.
While an analytical expectation of the lower limit of $P_1/2P_2$ is not available for cylinders,
    our study of an extremely large density contrast ($v_{\rm Ae}/v_{{\rm A}0}=20$) represented by the dashed curves in Fig.\ref{fig_Cor_PrSsg}
    shows that the sausage modes in a cylindrical geometry follow a similar pattern:
    $P_1/2P_2$ is also subject to a lower limit of $1/2$,
    the cutoff aspect ratio decreases with $v_{\rm Ae}/v_{{\rm A}0}$.
Likewise, the influence of flow on the standing modes is qualitatively similar in both geometries:
    introducing a flow in the structure has a more prominent effect in determining the cutoff aspect ratio
    than on the value of the period ratio.
With $M_{\rm A}$ in the examined range $[0, 0.6]$, in both geometries a flow may
    alter $(a/L)_{\mathrm{cutoff}}$ in an order-of-magnitude sense and the
    fractional change is more pronounced at higher density contrasts;
    whereas the fractional change in $P_1/2P_2$ with respect to the static case is $\lesssim 5\%$.

Figure~\ref{fig_Cor_PrSsg} also allows us to pay a closer inspection of the observed period ratios $P_1/2P_2$
    of standing sausage modes.
While $1-P_1/2P_2$ of standing kink modes has been examined in considerable
    detail~\citep[see e.g., the introduction in][and references therein]{2011A&A...526A..75M}
    and put in seismological applications~\citep[e.g.,][]{2005ApJ...624L..57A,2009SSRv..149....3A},
    the use of $1-P_1/2P_2$ of standing sausage modes seems not as popular~\citep[see e.g.,][]{2009A&A...503..569I}.
Before making its serious use, one may first ask the question that what leads to
    the departure of $P_1/2P_2$ from 1 in the first place.
The available data for $NoRH$ flare loops
    yield a value of
    $P_1/2P_2 \approx 15.5~{\rm s}/(2\times 9.5~{\rm s}) = 0.82$
    at an aspect ratio $a/L = 0.12$~\citep{2003A&A...412L...7N,2005A&A...439..727M},
    while those for cool H alpha post-flare loops yield a value of
    $P_1/2P_2 \approx 587~{\rm s}/(2\times 349~{\rm s}) = 0.84$
    at an $a/L=0.03$~\citep{2008MNRAS.388.1899S}.
In view of Fig.\ref{fig_Cor_PrSsg} which addresses trapped modes, the two values of $P_1/2P_2$ are difficult to explain:
    whichever value $a/L$ takes, $P_1/2P_2$ is far from the measured values, which are actually outside
    the range of the vertical extent of this figure.
Introducing a flow shear makes the comparison of the theoretically expected values with the measured ones
    even more undesirable: at a given $a/L$, $P_1/2P_2$ in the flowing case is actually even smaller
    than in the static case.
Adopting a slab description for coronal loops as was done in~\citet[][Fig.6]{2009A&A...503..569I}
    and in paper I (Figure 6 therein) does not help,
    varying the parameters of the equilibrium does not either,
    for the periods of standing modes are mostly determined by the density contrast~\citep{2009A&A...503..569I}.
On the other hand, in the leaky regime, the periods (and hence their ratios)
    subtly depend on the parameter range of the problem~\citep{2012ApJ...761..134N}.
We conclude that for sausage modes, what causes the deviation of $P_1/2P_2$ from unity
    needs a dedicated detailed investigation.

\begin{figure}
\centerline{\includegraphics[width=0.6\columnwidth]{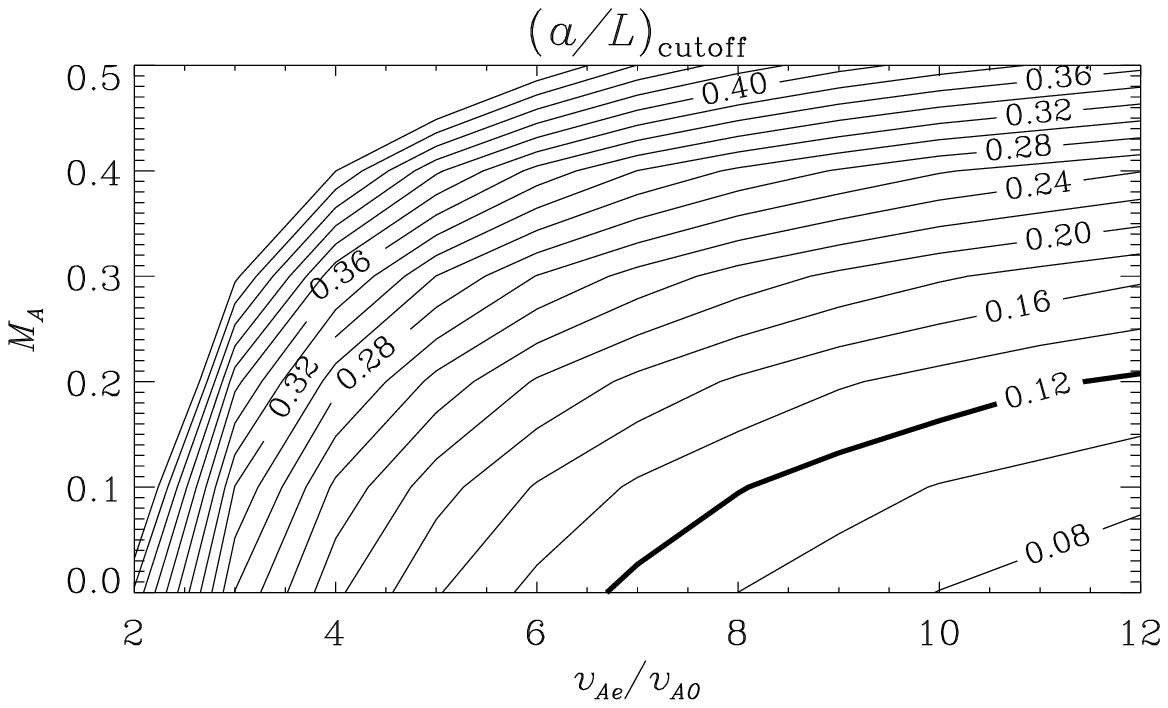}}
 \caption{The lowest allowed aspect ratio for standing sausage modes to occur, $(a/L)_{\mathrm{cutoff}}$,
    as a function of the Alfv\'en speed ratio $v_{\rm Ae}/v_{{\rm A}0}$ and the internal flow speed
    measured in units of the internal Alfv\'en speed.
The thick contour delineates where $(a/L)_{\mathrm{cutoff}} = 0.12$, corresponding to the aspect ratio
    of the flaring loop that experienced oscillations in the form of a global sausage mode as
    reported by~\citet{2003A&A...412L...7N}.
}
 \label{fig_Cor_Cutoff_contour}
\end{figure}
Figure~\ref{fig_Cor_Cutoff_contour} extends our examination on the effects of flow speed on
    the cutoff aspect ratio $(a/L)_{\mathrm{cutoff}}$ pertinent to the standing sausage modes by showing
    the distribution of $(a/L)_{\mathrm{cutoff}}$ with varying Alfv\'en speed ratios $v_{\rm Ae}/v_{{\rm A}0}$
    and Alfv\'en Mach numbers $M_{\rm A}$.
The contours of $(a/L)_{\mathrm{cutoff}}$ are equally spaced by $0.02$.
It can be seen that $(a/L)_{\mathrm{cutoff}}$ decreases monotonically with increasing $v_{\rm Ae}/v_{{\rm A}0}$ at a given $M_{\rm A}$,
    but increases rather dramatically with increasing $M_{\rm A}$ at a given $v_{\rm Ae}/v_{{\rm A}0}$.
What is more important in the context of SMS is that Fig.\ref{fig_Cor_Cutoff_contour} helps constrain the combinations of density contrast and
    internal flow, when trapped standing sausage modes are observed in a coronal cylinder with known aspect ratio.
This point can be illustrated if one examines the flaring loop reported in~\citet{2003A&A...412L...7N}, which
     is $25$~Mm long and $6$~Mm in diameter, resulting in an aspect ratio of $a/L = 0.12$.
Now that the fundamental sausage mode occurred in this loop, its aspect ratio has to be larger than the cutoff value,
     meaning that the pair of density contrast and internal flow has to be located in the region
     below the thick contour in Fig.\ref{fig_Cor_Cutoff_contour} which corresponds to $0.12$.
If the density contrast, or equivalently $v_{\rm Ae}/v_{{\rm A}0}$ is known, then the internal flow $U_0$ as measured in
     terms of $M_{\rm A} = U_0/v_{{\rm A}0}$ has to be smaller than some critical value, which in this particular example
     reads $\sim 0.16$ if $v_{\rm Ae}/v_{{\rm A}0}=10$, and $\sim 0.21$ if $v_{\rm Ae}/v_{{\rm A}0}=12$.
If one can further find the flow speed $U_0$ via, say, Doppler shift measurements using coronal emission lines, then one
     can derive a lower limit of the internal Alfv\'en speed.
For instance, supposing $U_0$ to be 40\velunits, found for warm EUV loops~\citep{2002ApJ...567L..89W}, one finds that $v_{{\rm A}0}$
     should be larger than $\sim 200$\velunits\ if $v_{\rm Ae}/v_{{\rm A}0}$ is found to be $\sim 12$.
Obviously this practice of SMS makes more sense when the contours in the upper half of Fig.\ref{fig_Cor_Cutoff_contour} can be employed,
    otherwise the deduced lower limit of $v_{{\rm A}0}$ is subject to large uncertainties.
Despite this and the difficulties associated with inferring the density contrast
    as well as flow speeds in coronal loops (see section~3.5 in \citeauthor{2010LRSP....7....5R}~\citeyear{2010LRSP....7....5R}),
    Fig.\ref{fig_Cor_Cutoff_contour} offers a possibility of exploiting the measured sausage oscillations.

\section{Period ratios for standing modes supported by photospheric Cylinders}
\label{sec_Pho_Pratio}

\subsection{Overview of Photospheric Cylinder Dispersion Diagrams}
\label{sec_sub_Pho_DR}

In this case the ordering $v_{{\rm A}0}>c_{\rm e}>c_0>v_{\rm Ae}$ holds.
Similar to~\citet{2003SoPh..217..199T}, only an isolated cylinder embedded
   in an unmagnetized atmosphere is considered:
   $v_{{\rm A}0} = 1.5 c_0, c_{\rm e} = 1.2 c_0, v_{\rm Ae}=0$ (and hence $c_{{\rm T}0} = 0.83 c_0, c_{{\rm Te}} = 0, \rho_{\rm e}/\rho_0 = 2$).
If assuming $c_0$ to be $8$\velunits, then one finds that $v_{{\rm A}0} = 12$ and
   $c_{\rm e} = 9.6$\velunits, which fall in category (ii) in~\citet{1990ApJ...348..346E}.
\begin{figure}
\centerline{\includegraphics[width=0.6\columnwidth]{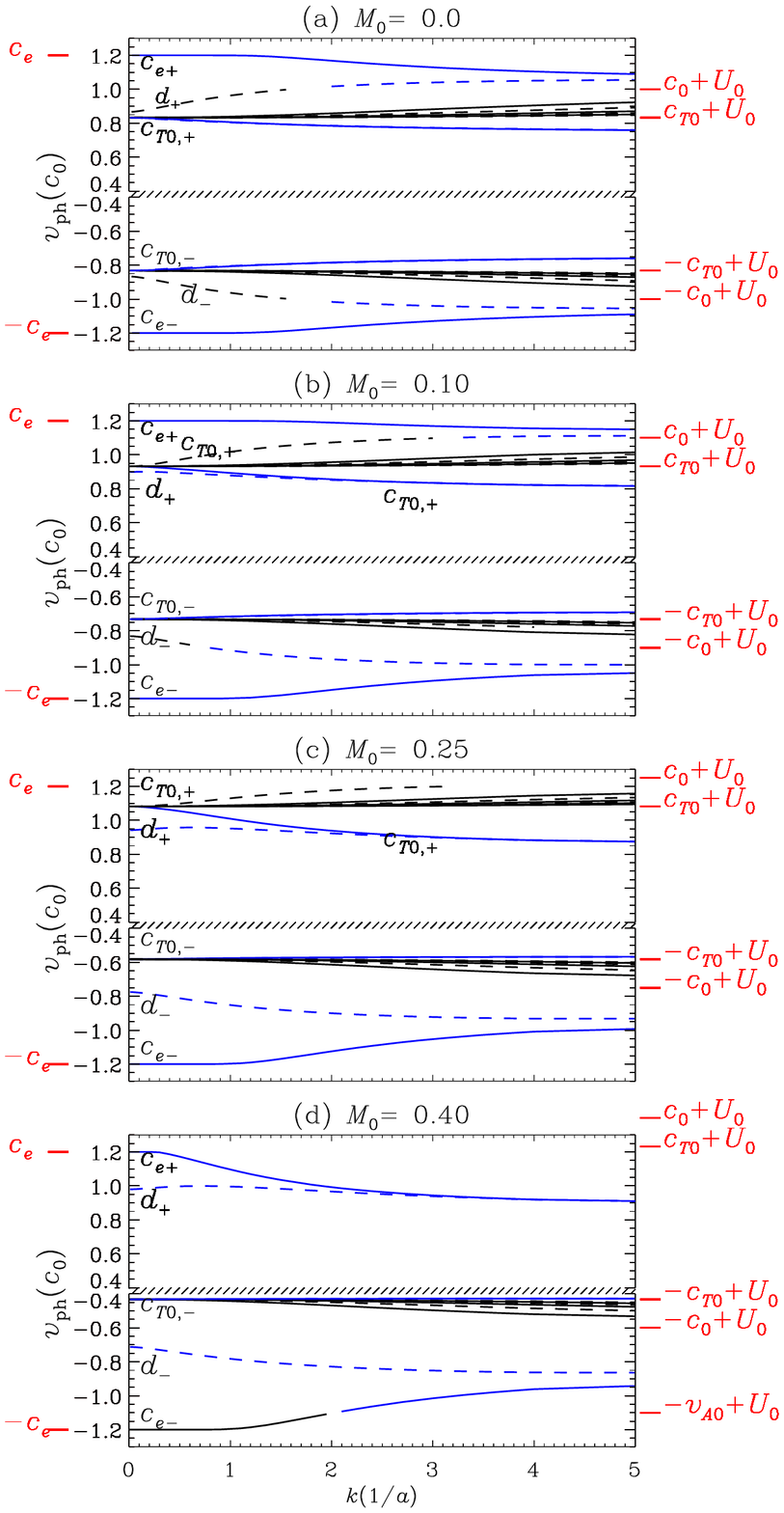}}
\caption{Similar to Fig.\ref{fig_Cor_DR} but for an isolated photospheric cylinder ($v_{\rm Ae} =0$).
The characteristic speeds are $c_{\rm e}=1.2 c_0, v_{{\rm A}0}=1.5 c_0$ and $c_{{\rm T}0}=0.83 c_0$.
Panels (a) to (d) correspond to an $M_0$ of $0, 0.1, 0.25$ and $0.4$, respectively, where
   $M_0 = U_0/c_0$ measures the internal flow in units of the internal sound speed.
The curves in blue (black) correspond to surface (body) waves.
The waves of interest are labeled with their phase speeds at zero longitudinal wavenumber $ka$, where
    $c_{{\rm e}\pm} = \pm c_{\rm e}$, and $c_{{\rm T}0,\pm} = \pm c_{{\rm T}0}+U_0$.
}
 \label{fig_Pho_DR}
\end{figure}

Figure~\ref{fig_Pho_DR} shows the dependence on longitudinal wavenumber $k$ of phase speeds $\vph$
   for a series of $U_0$, the magnitude of which is indicated by the internal Mach number $M_0 = U_0/c_0$.
Besides, the dashed (solid) curves are for kink (sausage) waves.
Note that the hatched area, corresponding to where $|\vph|\le 0.4$,
   does not contain any solutions to the DR, and therefore its vertical extent is artificially
   reduced to emphasize the area where solutions to the DR exist.
Labeling different waves is not as straightforward as in the coronal case,
   the reason being that in addition to body waves (the curves in black),
   surface waves are also allowed now (the curves in blue).
Instead of using the convention of grouping the wave modes into fast and slow ones~\citep{1990ApJ...348..346E},
   let us name them after their phase speeds at $ka=0$, with the exception
   of the majority of the body waves whose phase speeds are consistently
   bordered either by $c_{{\rm T}0}+U_0$ and $c_0+U_0$, or by $-c_0+U_0$ and $-c_{{\rm T}0}+U_0$.
This naming practice is necessary due to the change of identities of a number of wave modes in the presence of flow to be detailed shortly.
Note further that the surface waves labeled $c_{{\rm T}0,-}$ in all panels actually contain both
   a kink and a sausage solution, which can hardly be told apart though.
The same is also true for the $c_{{\rm T}0,+}$ surface waves in (a).
However, with increasing magnitude of $U_0$, the $c_{{\rm T}0,+}$ kink surface wave is replaced by
   the $d_{+}$ one, which becomes increasingly separated from the $c_{{\rm T}0,+}$ sausage surface wave
   (panels (b) and (c)).

While modest in magnitude, the flow has the subtle effect to make some propagating windows disappear
   as indicated by Fig.\ref{fig_Pho_DR}.
This is best illustrated by the body waves, which correspond to the two bands shifted upwards with increasing $U_0$.
For forward (backward) ones, and for both kink and sausage waves, it turns out in the slender
    cylinder limit ($ka \ll 1$)
    the behavior of the phase speeds $\vph$ can still be
    described by Eq.(\ref{eq_Cor_slow_smallk}).
Besides, in the opposite limit ($ka \gg 1$), with the exception of the $c_{{\rm T}0, +}$ wave
    the phase speeds $\vph$ approach $\pm c_0 + U_0$ in the same way as
    given by Eq.(\ref{eq_Cor_slow_bigk}).
The consequence is that, as trapped waves are bounded from above by $c_{\rm e}$,
    the propagation windows in the upper half-plane will disappear
    when $U_0 > c_{\rm e} - c_{{\rm T}0}$ (Fig.\ref{fig_Pho_DR}d).
It is also interesting to note that with varying $U_0$, the identity of wave modes may change.
For instance, in Fig.\ref{fig_Pho_DR}b while the $c_{{\rm T}0, +}$ wave starts at small $k$ as a body wave, it switches to
    a surface wave when $ka$ exceeds $\sim 3$ where $\vph$ exceeds $c_0 + U_0$.
This behavior was termed ``mode crossing'' as was noted in~\citet{2003SoPh..217..199T}.
Another feature is that, when $c_{\rm e} - c_{{\rm T}0} > U_0 > c_{\rm e} - c_0$ (Fig.\ref{fig_Pho_DR}c),
    the slow body kink waves can no longer be trapped when $ka$ exceeds
    a certain value, meaning that short wavelength waves then become leaky.

Other kink waves that undergo mode crossing include the $d_{+}$ mode in Fig.\ref{fig_Pho_DR}a, and
    the $d_{-}$ modes in Figs.\ref{fig_Pho_DR}a and \ref{fig_Pho_DR}b.
In the absence of flow, the $d_{\pm}$ modes are just the usual $c_k$ body modes,
    as examined extensively~\citep[e.g.,][]{1992SoPh..138..233G}.
At relatively low values of $U_0$, $d_{+}$ turns out to be smaller than $c_0 + U_0$,
    and $d_{-}$ larger than $-c_0 +U_0$ in an algebraic sense.
Now that $\vph$ increases (decreases) with increasing $ka$ for the upper (lower) branch,
    it eventually overtakes $c_0 + U_0$ ($-c_0 +U_0$), making the waves transition to
    surface ones.
Furthermore, as has been mentioned, when $U_0$ exceeds a certain value, $0.051 c_0$ to be specific,
    the kink body wave starts with $c_{{\rm T}0,+}$ instead, and likewise,
    the kink surface wave derives its label from $d_{+}$.
Of course, this particular value of $U_0$ is what makes $d_{+}$ equal to $c_{{\rm T}0,+}$.
When $ka \ll 1$, it turns out that be it a surface or a body wave, the $d_{\pm}$ waves have a phase speed that can still be
    approximated by Eq.(\ref{eq_Cor_FstKnk_smallK}),
    in which $\lambda_{\pm}$ may be simplified to $\lambda_{\pm} = \sqrt{1-d_{\pm}^2/c_{\rm e}^2}$ given that $v_{\rm Ae} = c_{{\rm Te}} = 0$.
Taking $U_0=0$, one recovers the static expression (11) in ER83 for photospheric cases.

Now move on to surface waves.
Consider first the $c_{{\rm T}0,\pm}$ ones.
It turns out that at $ka \ll 1$, the phase speeds of the kink ones labeled with $c_{{\rm T}0,\pm}$ have the form
\begin{eqnarray}
 \vph\ = U_0 \pm c_{{\rm T}0}\sqrt{1-\frac{c_{{\rm T}0}^2}{c_0^2 + v_{{\rm A}0}^2}\frac{(ka)^2}{\xi_{\pm}^2}} ,
\end{eqnarray}
    where $\xi_{\pm}$ are the solutions to the equation
\begin{eqnarray}
 \frac{x I_1'(x)}{I_1(x)} = \frac{c_{{\rm T}0}^2}{c_{{\rm T}0,\pm}^2}\frac{v_{{\rm A}0}^2}{c_0^2}\frac{\rho_0}{\rho_{\rm e}} ,
\end{eqnarray}
    with $x$ denoting the unknown.
This equation offers an extension to its static counterpart, Eq.(12), in ER83.
One may readily verify this by restricting oneself to the plus version, and by noting that
    $x I_1'(x)/I_1(x) = x I_0(x)/I_1(x)-1$.
Note that, when $U_0 > 0.051 c_0$, this transcendental equation has no solution when the plus sign is adopted,
    for beyond this $U_0$ the kink surface waves start with $d_{+}$ when $ka \rightarrow 0$.
On the other hand, for the sausage ones labeled $c_{{\rm T}0,\pm}$, one has for $ka \ll 1$
\begin{eqnarray}
 \vph = U_0 \pm c_{{\rm T}0}\sqrt{1-\frac{1}{2}\frac{\rho_{\rm e}}{\rho_0}\frac{c_{{\rm T}0, \pm}^2 c_{{\rm T}0}^2}{v_{{\rm A}0}^4}\left(ka\right)^2 K_0(\chi_{\pm} ka)}
\end{eqnarray}
where $\chi_{\pm} = \sqrt{1-c_{{\rm T}0,\pm}^2/c_{\rm e}^2}$.
This equation agrees closely with Equation (27) in TEB03, save the typo therein that the parentheses in the first line
    should be removed.
Despite the difference in the form of $\vph$ for the kink and sausage waves, one can hardly discern
    the difference between the kink and sausage ones sharing the label $c_{{\rm T}0,-}$ in Figs.\ref{fig_Pho_DR}a to \ref{fig_Pho_DR}d.
For the $c_{{\rm T}0,+}$ sausage one, while it virtually merges with the $d_{+}$ kink one at sufficiently large wavenumbers,
    which reads $ka \sim 1.8$ in Fig.\ref{fig_Pho_DR}b,
    its difference from the $d_{+}$ kink one becomes more and more obvious with
    increasing $U_0$ at small $ka$.
Now consider the $c_{{\rm e} \pm}$ surface ones, where $c_{{\rm e} \pm} = \pm c_{\rm e}$.
One can see that while in all panels the $c_{{\rm e}-}$ wave exists, it is slightly different in panel (d) where
    it is a body wave at $ka \lesssim 2$.
This is understandable because at this $U_0$ the Doppler-shifted Alfv\'en speed $-v_{{\rm A}0}+U_0$ is actually
    larger than $-c_{\rm e}$, thereby making $m_0^2$ negative at small $k$
    (see Eq.(\ref{eq_def_m2})).
Concerning the $c_{{\rm e}+}$ mode, it tends asymptotically at $ka \gg 1$ to some value slightly above $c_0+U_0$,
    as shown in Figs.\ref{fig_Pho_DR}a and \ref{fig_Pho_DR}b.
As such, when $U_0 > c_{\rm e}-c_0$, this mode disappears as shown in Fig.\ref{fig_Pho_DR}c.
At sufficiently strong $U_0> c_{\rm e} - c_{{\rm T}0}$, the $c_{{\rm T}0,+}$ sausage surface mode starts with $c_{\rm e}$ instead.
One can see that only sausage solutions are allowed, and these are not subject to a cutoff wavenumber at small $k$,
    meaning that
    even thin cylinders with tiny aspect ratios can support standing sausage modes that are formed by
    a pair of $c_{{\rm e}+}$ and $c_{{\rm e}-}$ propagating waves.
It is informative to consider analytically the nearly dispersionless range of $ka$ where $\vph$ is literally $\pm c_{\rm e}$.
If this range is not considered part of the solution to the DR, then one may have the impression that a low-wavenumber cutoff exists,
    which actually is not the case.
It turns out that $\vph$ at small $ka$
    can be approximated by
\begin{eqnarray}
 \vph \approx \pm c_{\rm e} \sqrt{1-\frac{4}{k^2 a^2}\exp\left[-\frac{\eta_\pm}{(ka)^2}\right]} ,
\label{eq_vph_sausage_ce}
\end{eqnarray}
    where
\begin{eqnarray}
 \eta_\pm = \frac{4 \rho_0}{\rho_{\rm e}}\frac{(c_0^2 + v_{{\rm A}0}^2)(\bar{c}_{{\rm e},\pm}^2 - c_{{\rm T}0}^2)}{c_{\rm e}^2 (\bar{c}_{{\rm e},\pm}^2 - c_{0}^2)},
\end{eqnarray}
    with $\bar{c}_{{\rm e},\pm} = \pm c_{\rm e} -U_0$.
Previous studies correctly suspected that this apparent cutoff may be caused by the difficulty for a numerical DR solver
    to resolve adequately the difference between $\vph$ and $\pm c_{\rm e}$ at small wavenumbers
    (see \citeauthor{2013A&A...551A.137M}~\citeyear{{2013A&A...551A.137M}},
    \citeauthor{2010SoPh..263...63E}~\citeyear{2010SoPh..263...63E}).
Equation~(\ref{eq_vph_sausage_ce}) shows that the particular $ka$ dependence of $\vph$ is the
    culprit for this numerical difficulty.
The same approximate expression also applies to the $c_{{\rm e}-}$ mode in Fig.\ref{fig_Pho_DR}d even though it starts as a body wave.

\subsection{Computing Standing Modes}
\label{sec_sub_comp_pho_pratios}

Constructing standing modes requires us to properly choose a pair of propagating waves,
    and by saying a combination is realistic or not we mean
    the resulting standing mode corresponds to a realistic density fluctuation in the slender cylinder limit.
Let us recall that for the two propagating waves in question,
    they may be both kink ones or sausage ones, but are not allowed to be a mixture of the two kinds.

Let us show that the combinations involving one or two slow body waves are not of interest.
Here ``slow body waves'' refer to the body waves that have phase speeds close to $\pm c_{{\rm T}0} + U_0$ at
    small wavenumbers with the exception of those labeled $d_{+}$ and $c_{{\rm T}0,+}$ in Fig.\ref{fig_Pho_DR}.
The combination of a forward and backward slow body wave may be interesting in its own right,
    but is not so when the period ratios are concerned in view of the very
    mild dispersion these waves possess.
Is it then possible that one of the two propagating waves is a slow body one,
    but the other is not?
Once again, this turns out unlikely because when $ka \rightarrow 0$, for slow body waves $X$ tends to large values for the kink and sausage waves alike,
    whereas for all other wave modes $X$ either tends to zero or to something finite.
The reason is given as follows.

For the ease of discussion, let us rewrite Eqs.(\ref{eq_Bdy_rhoOxi}) and (\ref{eq_Srfc_rho0xi}) as
\begin{eqnarray}
 X = \Lambda_v \Lambda_x ,
\end{eqnarray}
    where
\begin{align*}
& \Lambda_v = (\vph - U_0)^2/[c_0^2 -(\vph - U_0)^2], \\
& \Lambda_x = \left\{
  \begin{array}{l l}
      -(n_0 a) J_n(n_0 a)/J_n'(n_0 a)   \quad \text{, for kink ($n=1$) and sausage ($n=0$) body waves,} \\
       (m_0 a) I_n(m_0 a)/I_n'(m_0 a)   \quad \text{, for kink ($n=1$) and sausage ($n=0$) surface waves.}
  \end{array} \right.
\end{align*}
For waves with phase speeds starting with $c_{{\rm T}0, \pm}$, one finds $\Lambda_v = v_{{\rm A}0}^2/c_0^2$, which evaluates to $2.25$.
Also of interest are the waves with $\vph$ starting with $c_{{\rm e}\pm}$, for which $\Lambda_v$ lies in the interval between
    $-9.26$ and $5.76$.
Furthermore, for slow kink body waves, with $ka$ approaching zero, $n_0 a$ approaches $h_l$ given by Eq.(\ref{eq_hj_KINK}).
For the photospheric computations $h_l$ is found to be rather well approximated by $(l+3/4)\pi$, making $J_1'(n_0 a)$ tend to zero
     and hence $\Lambda_x$ approach big values.
Likewise, for slow sausage body ones, when $ka$ tends to $0$, $n_0 a$ as given by $h_l$ in Eq.(\ref{eq_hj_sausage})
    causes $J_0'(n_0 a)$ to approach zero and consequently $\Lambda_x$ to tend to infinity.
Now to examine the rest of the labeled waves, we may start with the kink category.
It can be shown that the $d_{\pm}$ waves, be them body or surface ones,
    correspond to an $X$ that tends to zero when $ka\rightarrow 0$, because
    $\Lambda_x \rightarrow (m_0 a)^2$.
Furthermore, for the $c_{{\rm T}0,+}$ kink body wave in Figs.\ref{fig_Pho_DR}b to~\ref{fig_Pho_DR}d, one may find that
    $\Lambda_x$ in the zero wavenumber limit ranges from $-3.5$ to $-0.45$.
In Fig.\ref{fig_Pho_DR}a this $c_{{\rm T}0,+}$ kink mode is a sausage one and one finds $\Lambda_x = 0.46$.
Its minus counterpart, the $c_{{\rm T}0, -}$ kink mode, is always a surface one and one finds $\Lambda_x$ ranges between
    $0.46$ and $5$.
On the other hand, the $c_{{\rm T}0,\pm}$ sausage surface waves correspond to $\Lambda_x = 2$ when $ka \rightarrow 0$.
The same value of $\Lambda_x$ at zero $ka$ is found for all the $c_{{\rm e}\pm}$ sausage waves,
    regardless of whether they belong to the body or surface category.

\subsection{Period Ratios for Standing Kink Modes}
\label{sec_sub_Pho_PrKnk}

\begin{figure}
\centerline{\includegraphics[width=0.6\columnwidth]{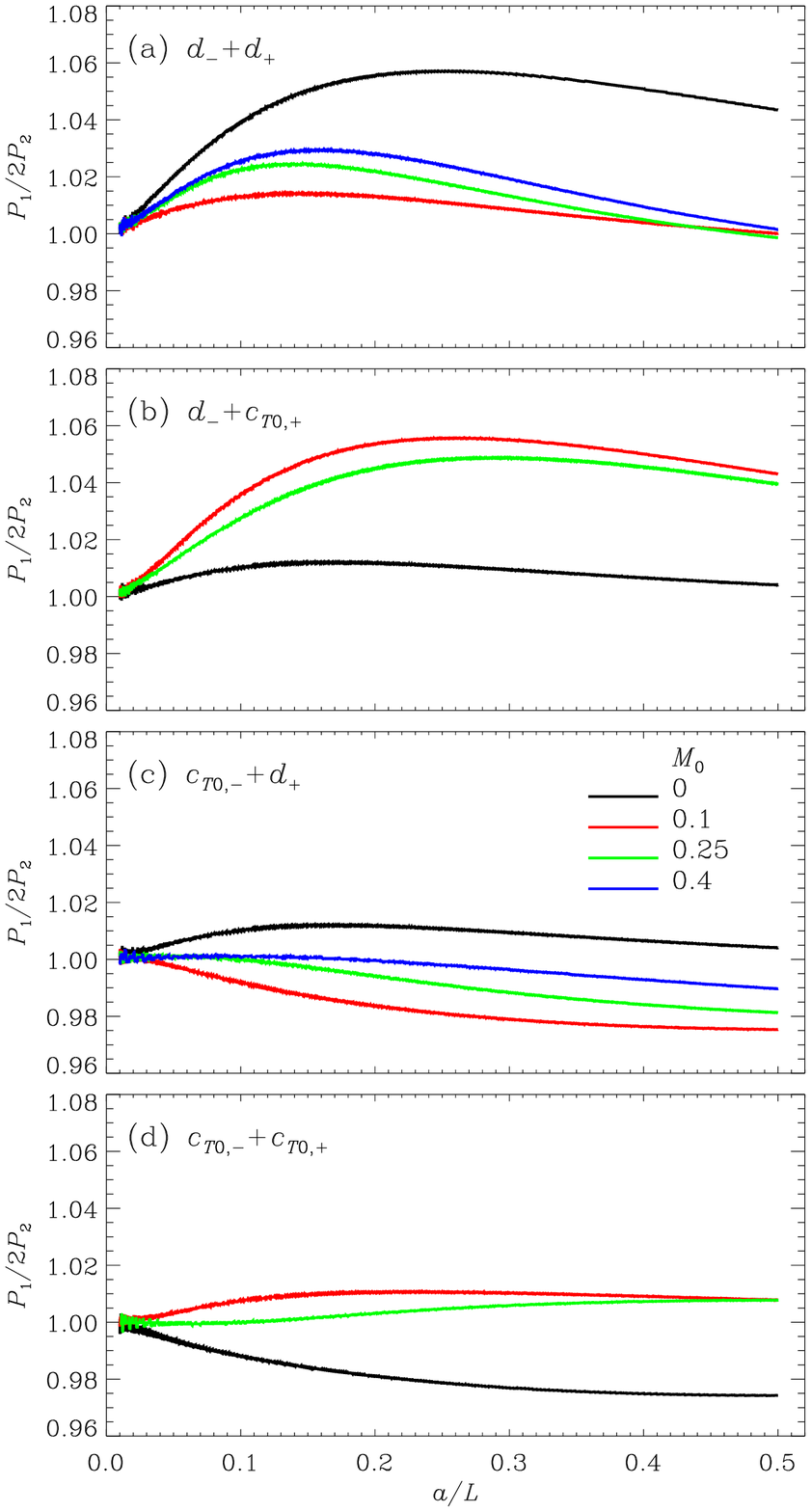}}
 \caption{Period ratio $P_1/2P_2$ as a function of the cylinder aspect
   ratio $a/L$ for standing kink modes.
Curves with different colors represent results computed for different values of the
   flow speed $U_0$ measured in units of the internal sound speed $U_0 = M_0 c_{0}$.
Presented in (a) to (d) are combinations of $d_{-}+d_{+}$, $d_{-}+c_{{\rm T}0,+}$,
   $c_{{\rm T}0,-}+d_{+}$, and $c_{{\rm T}0,-}+c_{{\rm T}0,+}$, respectively.
}
 \label{fig_Pho_PrKnk}
\end{figure}

Figure~\ref{fig_Pho_PrKnk} presents the aspect ratio dependence of the period ratio $P_1/2P_2$ for standing kink modes
   for a number of $U_0$ indicated by different colors.
Distinct from the coronal case, to construct standing modes, one is allowed to pick one component from
   the $d_{\pm}$ waves, and the other from the $c_{{\rm T}0, \pm}$ ones, resulting in four possible combinations, namely,
   ``$d_{-} + d_{+}$'', ``$d_{-} + c_{{\rm T}0, +}$'', ``$c_{{\rm T}0, -} + d_{+}$'', and ``$c_{{\rm T}0, -} + c_{{\rm T}0, +}$''.
Note that instead of four curves, there are only three in Figs.\ref{fig_Pho_PrKnk}b and \ref{fig_Pho_PrKnk}d,
    since the $c_{{\rm T}0,+}$ wave does not exist
    when $M_0=0.4$.
Besides, as opposed to the coronal case where the period ratios are consistently less than one,
    now $P_1/2P_2$ may be larger than $1$, as seen in the combinations involving $d_{-}$ (Figs.\ref{fig_Pho_PrKnk}a and~\ref{fig_Pho_PrKnk}b),
    as well as the ``$c_{{\rm T}0,-} + d_{+}$'' one in the static case (black curve in Fig.\ref{fig_Pho_PrKnk}c),
    and the ``$c_{{\rm T}0,-} + c_{{\rm T}0, +}$'' one when $M_0$ is $0.1$ or $0.25$ (red and green curves in Fig.\ref{fig_Pho_PrKnk}d).
Evidently, this happens when one or both waves in a combination corresponds to a phase speed that increases in magnitude
   with increasing wavenumber in part of or the whole range of considered wavenumbers.

Overall, the flow effect is rather modest in the parameter range explored.
Consider Fig.\ref{fig_Pho_PrKnk}a for instance, where the flow effect is almost the strongest in the four.
One can see that the maximum the period ratio attains, $(P_1/2P_2)_{\mathrm {max}}$, reads $1.057$ in the static case.
The largest deviation from this occurs when $U_0 = 0.1 c_0$,
    where $(P_1/2P_2)_{\mathrm {max}}$ reads $1.014$, resulting in a fractional difference of $4.1\%$.
That this is not associated with the largest flow speed results from the fact that
   when $U_0$ exceeds $0.051 c_0$,
   the $c_{{\rm T}+}$ mode is shifted upwards to an extent that it takes the original position of the $d_{+}$ mode,
   as discussed regarding Fig.\ref{fig_Pho_DR}.
Consequently, beyond this particular value $P_1/2P_2$ tends to increase rather than decrease with increasing $U_0$.
The same $U_0$ dependence of the period ratio also occurs in the rest of the panels.
For instance, for the combination ``$d_{-} + c_{{\rm T}0, +}$'', one can see from Fig.\ref{fig_Pho_PrKnk}b that $(P_1/2P_2)_{\mathrm {max}}$
   reads $1.012$ in the static case, and increases to $1.056$ when $M_0=0.1$, corresponding to a relative difference of $4.3\%$.
For the combinations involving $c_{{\rm T}0, -}$, Figs.\ref{fig_Pho_PrKnk}c and~\ref{fig_Pho_PrKnk}d indicate that the flow effect is less pronounced.
In the case of ``$c_{{\rm T}0, -}+d_{+}$'' (``$c_{{\rm T}0, -}+d_{+}$''), the maximal relative difference in the extremes of $P_1/2P_2$
   reads $3.7\%$ ($3.8\%$).

\subsection{Period Ratios for Standing Sausage Modes}
\label{sec_sub_Pho_PrSsg}

\begin{figure}
\centerline{\includegraphics[width=0.6\columnwidth]{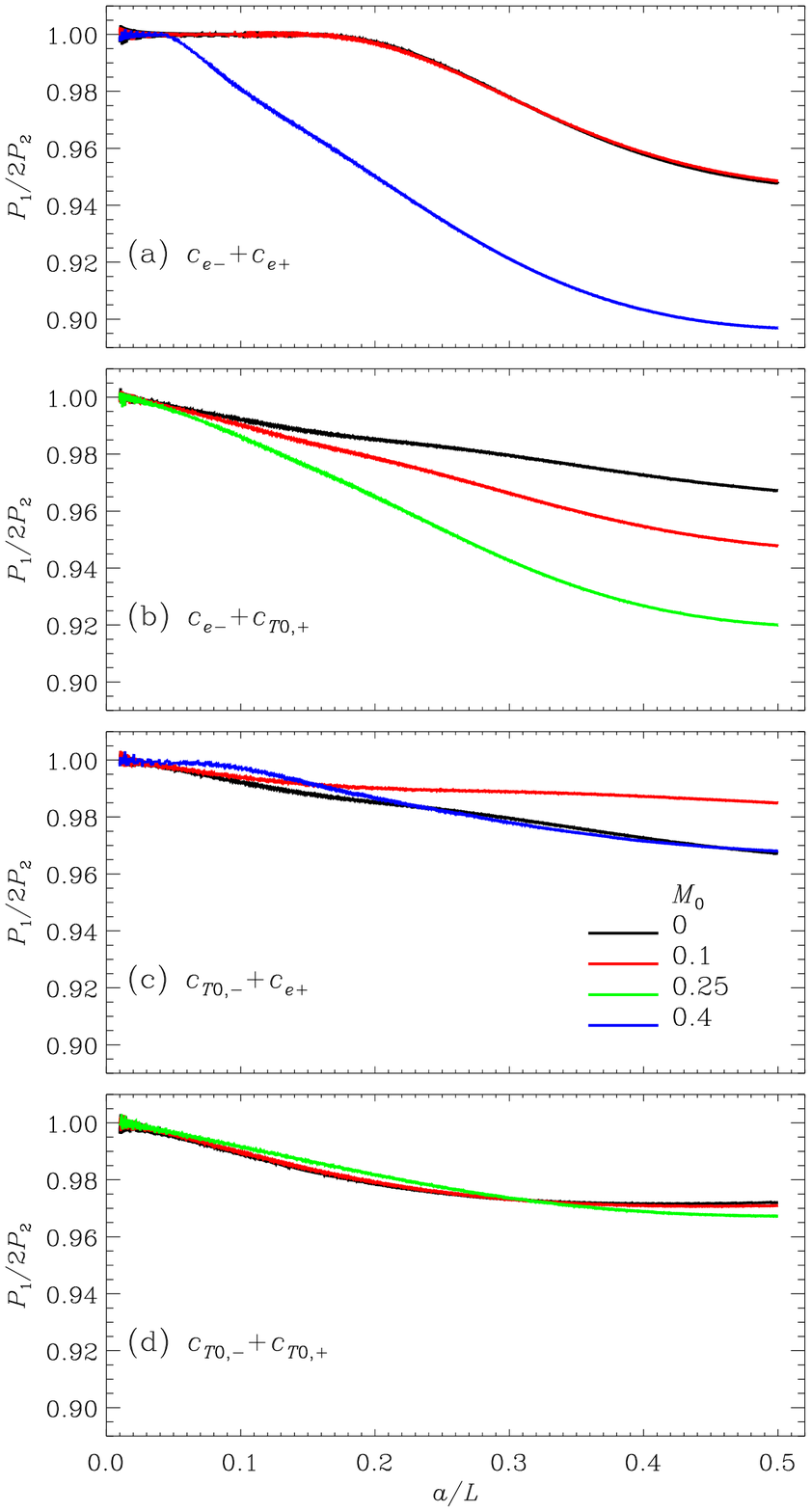}}
 \caption{Similar to Fig.\ref{fig_Pho_PrKnk} but for standing sausage modes.
Presented in (a) to (d) are combinations of $c_{{\rm e}-}+c_{{\rm e}+}$, $c_{{\rm e}-}+c_{{\rm T}0,+}$,
   $c_{{\rm T}0,-}+c_{{\rm e}+}$, and $c_{{\rm T}0,-}+c_{{\rm T}0,+}$, respectively.
}
 \label{fig_Pho_PrSsg}
 \end{figure}

Figure~\ref{fig_Pho_PrSsg} presents the aspect ratio dependence of the period ratio $P_1/2P_2$
   for a number of $U_0$ pertinent to standing sausage modes.
As opposed to the coronal case, one can see no cutoff in $a/L$ any longer, meaning that
   cylinders with arbitrary aspect ratios can support sausage modes.
Moreover, all curves start with unity at zero aspect ratio, indicating that wave dispersion is negligible at small wavenumbers
   for any component wave that is employed to construct a standing mode.
Four pairs of combinations,
   ``$c_{{\rm e}-}+c_{{\rm e}+}$'', ``$c_{{\rm e}-}+c_{{\rm T}0,+}$'', ``$c_{{\rm T}0,-}+c_{{\rm e}+}$'', and ``$c_{{\rm T}0,-}+c_{{\rm T}0,+}$''
   are possible and presented from top to bottom, respectively.
Note that when $M_0$ is $0.25$, the $c_{{\rm e}+}$ mode does not exist, hence there are
   only three curves in Figs.\ref{fig_Pho_PrSsg}a and~\ref{fig_Pho_PrSsg}c.
Likewise, there are no blue curves in Figs.\ref{fig_Pho_PrSsg}b and~\ref{fig_Pho_PrSsg}d,
   since the $c_{{\rm T}0, +}$ wave is absent
   in the $M_0=0.4$ case.

The flow effect is stronger than for the standing kink modes as far as the period ratio is concerned.
While hardly discernible for the combination ``$c_{{\rm T}0,-}+c_{{\rm T}0,+}$'' (Fig.\ref{fig_Pho_PrSsg}d)
   and marginal for ``$c_{{\rm T}0,-}+c_{{\rm e}+}$'' (Fig.\ref{fig_Pho_PrSsg}c),
   the flow effect is rather significant for the top two combinations.
In the case of ``$c_{{\rm T}0,-}+c_{{\rm e}+}$'', from Fig.\ref{fig_Pho_PrSsg}b one can see that
   introducing a finite $U_0$ leads to a decrease in $P_1/2P_2$ in general.
The relative change between the static case and the one with $M_0=0.25$ may reach $4.89\%$,
   attained at the biggest aspect ratio considered.
In the case of ``$c_{{\rm e}-} + c_{{\rm e}+}$'',
    one can see from Fig.\ref{fig_Pho_PrSsg}a that while there is no difference between the two curves corresponding the static case
    and the case where $M_0=0.1$,
    significant changes appear for aspect ratios above $\sim 0.05$ when $U_0$ is further increased to $0.4 c_0$.
With this $M_0$, $P_1/2P_2$ may decrease by up to $5.85\%$ relative to the static case.
Note that even though the fractional change in $P_1/2P_2$ of a few percent is similar to what was found for standing kink modes,
    the deviation of $P_1/2P_2$ from unity is considerably more prominent, with
    $P_1/2P_2$ reaching as low as $0.897$ when $M_0=0.4$.
For comparison, $|P_1/2P_2-1|$ for standing kink modes always lies below $0.06$ regardless of the combinations.

Our examination of standing sausage modes may be relevant for interpreting
    the very recent direct measurements of sausage oscillations with imaging instruments~\citep{2011ApJ...729L..18M}
    (hereafter MEJM, see also \citeauthor{2008IAUS..247..351D}~\citeyear{2008IAUS..247..351D}, \citeauthor{2012NatCom..3.1315M}~\citeyear{2012NatCom..3.1315M}).
Let us focus on the multiple periods revealed therein.
Let $P_1$ stand for the period of the fundamental, and $P_n$ ($n=2, 3, ...$) denote
    that of the $(n-1)$-th overtone.
For illustrative purpose, restrict ourselves to $\hat{P}=281\pm 18$~s.
Taking $P_1$ to be $550$~s pertinent to the fast modes (the distinction between fast and slow modes
    is described in quite some detail by~\citeauthor{1990ApJ...348..346E}~\citeyear{1990ApJ...348..346E}), one finds that
    $P_1/nP_n$ would lie in the range $[0.92, 1.05]$ ($[0.61, 0.70]$)
    if $\hat{P}$ corresponds to the first (second) overtone.
Now that the loop aspect ratio in question was measured to be $a/L$ is $\sim 0.3$,
    our Figs.\ref{fig_Pho_PrSsg}a and \ref{fig_Pho_PrSsg}b indicate that
    the computed $P_1/2P_2$ lies in the range that corresponds to $n=2$.
This lends support to the suggestion by MEJM that this $\hat{P}$ corresponds to the first overtone.
Is it possible to be related to the second overtone?
This turns out to be unreasonable since a computation yields that $P_1/3P_3$ for the combination ``$c_{{\rm e}-}+c_{{\rm e}+}$'' cannot
    drop below $0.86$.
If one chooses $P_1$ to be $660$~s pertinent to slow modes instead,
   then $\hat{P}$ would correspond to a $P_1/nP_n$ in the range $[1.1, 1.26]$ ($[0.74, 0.84]$)
   if it corresponds to $n=2$ ($n=3$).
The former can be ruled out, since we have seen that for photospheric standing sausage modes,
   $P_1/2P_2$ never exceeds unity.
The latter does not appear to be likely either.
This is because, strictly speaking, by convention slow modes correspond to the combination ``$c_{{\rm T}0,-}+c_{{\rm T}0,+}$'',
   for which we find that $P_1/3P_3$ is in excess of $0.954$, i.e., outside
   the deduced range for all the flow speeds considered.
Despite that this comparison is admittedly inconclusive,
   let us make the point that the incorporation of flow shear in addition to a transverse density structuring
   offers more possibilities in interpreting the measured oscillation periods.

\section{Summary and Concluding Remarks}
\label{sec_summary}

The present study is dedicated to examining the effects of a field-aligned flow
    on the period ratios $P_1/2P_2$, where
    $P_1$ and $P_2$ represent the periods of the fundamental and its first overtone,
    for both standing kink and sausage modes, and for both a coronal and a photospheric environment.
It was motivated by the fact that in the field of solar magneto-seismology (SMS),
    multiple periodicities are playing an increasingly important role on the one hand~\citep[e.g.,][]{2009SSRv..149....3A,2009SSRv..149..199R},
    significant flows were found to have important consequences for seismological applications
    on the other~\citep{2011ApJ...729L..22T}.
While our previous work~\citep{2013ApJ...767..169L} sees magnetic loops as slabs,
    here they are modeled as magnetized cylinders.
To be specific, we numerically solve the dispersion relations for waves supported by cylinders incorporating flows,
    devise a graphical method to construct standing kink and sausage modes,
    and examine in detail how the period ratios depend on the loop aspect ratio $a/L$,
    the flow magnitude, as well as the density contrast between the loop and its surroundings.
Here $a$ is the loop radius, and $L$ is its length.
Concerning the period ratios, our conclusions can be summarized as follows.

\begin{enumerate}
\item
For standing kink modes supported by coronal cylinders,
    introducing a significant field-aligned flow in the cylinder
    may reduce the period ratio by up to 17\% compared with the static case.
This fractional change depends only weakly on the density contrast,
    a similar amount of reduction is found even in the limit where the density contrast approaches infinity.
In addition, the reduction in the period ratio due to a finite flow may readily help explain
    the observed values at finite aspect ratios of the recently reported oscillating $NoRH$ loops,
    and is not negligible for thin cylinders at large shear flows (high $U_{0}$).
\item
For standing sausage modes supported by coronal cylinders,
    even a significant flow can only lead to a reduction in $P_1/2P_2$ that is typically no more than $5.5$\%
    relative to the static case.
Despite that, it has important effects on the threshold aspect ratio
    only above which standing sausage modes can be supported.
At a given density contrast, this threshold may be larger than its static counterpart by an order-of-magnitude.
On the one hand, this may explain why the measured standing sausage modes are rare since the existence of
    a flow in the loop makes the modes more difficult to be trapped.
On the other hand, we show that this parameter dependence of the threshold may be exploited to constrain
    the combinations of density contrast $\rho_0/\rho_{\rm e}$ and Alfv\'en Mach number $M_{\rm A}$.
If the density contrast and flow speed are further known, then this practise can help yield the lower limit of the internal
    Alfv\'en speed.

\item
For the isolated photospheric cylinders, we find that the flow effect is marginal
    on the period ratios $P_1/2P_2$ for the standing sausage modes,
    and even less so for the kink modes.
Having said that, we note that standing modes in this case are distinct from the coronal case in that
    standing sausage modes may be supported by cylinders with arbitrary aspect ratios and are not subject to an
    aspect ratio cutoff any more.
Furthermore, for standing kink modes $P_1/2P_2$ may exceed unity as a result of the wavenumber dependence of the phase speed.

\item
While this study focuses on the period ratios of standing modes, it offers some new results for the dispersion properties
    of propagating waves as well, in the form of a series
    of approximate expressions for the phase speed $\vph$ in both slender ($ka \ll 1$) and thick ($ka \gg 1$) cylinder limits.
In particular, the expression for $\vph$ in the slender cylinder limit for photospheric loops (Eq.\ref{eq_vph_sausage_ce})
    provides an explanation for the numerical difficulty associated with finding solutions to the dispersion relation
    pertinent to sausage waves in this limit.
\end{enumerate}

Before closing, a few remarks on the applications of the presented study are necessary.
First, let us stress that allowing the loop parameters to be time-dependent may be important
    as far as the period ratio is concerned~\citep{2009ApJ...707..750M,2013SoPh..283..413A,2013SoPh..tmp..195E},
    and hence it is necessary to address the consequence of a time-varying flow speed in this regard.
Our results on the effect of flow on standing sausage modes in a coronal environment makes such a further
    investigation particularly necessary.
Second, in agreement with~\citet{2011ApJ...729L..18M}, the idea of seismology may be equally applicable
    to other parts of the structured solar atmosphere,
    the photospheric structures in particular.

\acknowledgements
    This research is supported by the 973 program 2012CB825601, the National Natural Science Foundation of China
    (41174154, 41274176, 41274178, and 41474149),
    and by
    the Provincial Natural Science Foundation of Shandong via Grant JQ201212.







%

%

%

%


\end{document}